\documentclass[twocolumn]{aastex63}

\usepackage{scrextend}
\let\orgautoref\autoref
\renewcommand{\autoref}
        {\def\equationautorefname{Eq.}%
         \def\figureautorefname{Fig.}%
         \def\sectionautorefname{Sect.}%
         \def\subsectionautorefname{Sect.}%
         \def\subsubsectionautorefname{Sect.}%
         \orgautoref}

\shorttitle{Dynamical architecture of the HD\,107148 system}
\shortauthors{Eberhardt et al.}
\graphicspath{{./}{}}

\begin{document}

\title{Dynamical architecture of the HD\,107148 system}

\author[0000-0003-3130-2768]{Jan Eberhardt}
\affiliation{Max-Planck-Institut für Astronomie,
              Königstuhl 17,
              69117 Heidelberg, Germany}

\author[0000-0002-0236-775X]{Trifon Trifonov}
\affiliation{Max-Planck-Institut für Astronomie,
              Königstuhl 17,
              69117 Heidelberg, Germany}

\author[0000-0002-1765-9907]{Martin Kürster}
\affiliation{Max-Planck-Institut für Astronomie, Königstuhl 17, 69117 Heidelberg, Germany}

\author[0000-0002-1166-9338]{Stephan Stock}
\affiliation{Landessternwarte, Zentrum für Astronomie der Universtät Heidelberg,
              Königstuhl 12,
              69117 Heidelberg, Germany}

\author[0000-0002-1493-300X]{Thomas Henning}
\affiliation{Max-Planck-Institut für Astronomie,
              Königstuhl 17,
              69117 Heidelberg, Germany}

\author{Anna Wollbold}
\affiliation{Max-Planck-Institut für Astronomie,
              Königstuhl 17,
              69117 Heidelberg, Germany}

\author[0000-0002-0460-8289]{Sabine Reffert}
\affiliation{Landessternwarte, Zentrum für Astronomie der Universtät Heidelberg,
              Königstuhl 12,
              69117 Heidelberg, Germany}

\author[0000-0003-1930-5683]{Man Hoi Lee}
\affiliation{Department of Earth Sciences, The University of Hong Kong,
Pokfulam Road,
Hong Kong}
\affiliation{Department of Physics, The University of Hong Kong,
Pokfulam Road,
Hong Kong }

\author[0000-0002-6532-4378]{Mathias Zechmeister}
\affiliation{Institut für Astrophysik, Georg-August-Universität,
              Friedrich-Hund-Platz 1,
              37077 Göttingen, Germany}

\author[0000-0003-0650-5723]{Florian Rodler}
\affiliation{European Southern Observatory (ESO),
Alonso de Cordova 3107,
              Vitacura, Santiago de Chile, Chile}

\author[0000-0002-4199-6356]{Olga Zakhozhay}
\affiliation{Max-Planck-Institut für Astronomie,
              Königstuhl 17,
              69117 Heidelberg, Germany}
\affiliation{Main Astronomical Observatory, National Academy of Sciences of Ukraine, Kyiv 03143, Ukraine}

\author[0000-0002-3662-9930]{Paul Heeren}
\affiliation{Landessternwarte, Zentrum für Astronomie der Universtät Heidelberg,
              Königstuhl 12,
              69117 Heidelberg, Germany}

\author[0000-0001-8627-9628]{Davide Gandolfi}
\affiliation{Dipartimento di Fisica, Università di Torino,
via P. Giuria 1,
         10125 Torino, Italy}

\author[0000-0003-0563-0493]{Oscar Barragán}
\affiliation{Sub-department of Astrophysics, Department of Physics, University of Oxford, Oxford OX1 3RH, UK}

\author[0000-0003-1566-7740]{Marcelo Tala Pinto}
\affiliation{Facultad de Ingeniería y Ciencias, Universidad Adolfo Ibañez, Av. Diagonal las Torres 2640, Peñalolen, Santiago, Chile}

\author{Vera Wolthoff}
\affiliation{Landessternwarte, Zentrum für Astronomie der Universtät Heidelberg,
              Königstuhl 12,
              69117 Heidelberg, Germany}

\author[0000-0001-8128-3126]{Paula Sarkis}
\affiliation{Max-Planck-Institut für Astronomie,
              Königstuhl 17,
              69117 Heidelberg, Germany}

\author[0000-0002-1440-3666]{Stefan S. Brems}
\affiliation{Landessternwarte, Zentrum für Astronomie der Universtät Heidelberg,
              Königstuhl 12,
              69117 Heidelberg, Germany}

\begin{abstract}

We present an independent Doppler validation and dynamical orbital analysis of the two-planet system HD\,107148,
which was recently announced in \citet{Rosenthal2021}. Our detailed analyses are based on literature HIRES data and newly obtained HARPS and CARMENES radial velocity (RV) measurements as part of our survey in search for additional planets around single planet systems.
We perform a periodogram analysis of the available HIRES and HARPS precise RVs and stellar activity indicators. We do not find any apparent correlation between the RV measurements and the stellar activity indicators, thus linking the two strong periodicities to a moderately compact multiple-planet system.
We carry out orbital fitting analysis by testing various one- and two-planet orbital configurations and studying the posterior probability distribution of the fitted parameters.
Our results solidify the existence of a Saturn-mass planet (HD\,107148\,b, discovered first) with a period $P_b\sim77.2$\,d,
and a second, eccentric ($e_c \sim$ 0.4), Neptune-mass exoplanet  (HD\,107148\,c), with an orbital period of
$P_c\sim18.3$\,d.
Finally, we investigate the two-planet system's long-term stability and overall orbital dynamics with the posterior distribution of our preferred orbital configuration.
Our N-body stability simulations show that the system is long-term stable and exhibits large secular osculations in eccentricity but in no particular mean-motion resonance configuration.
The HD\,107148 system, consisting of a Solar-type main sequence star with two giant planets in a rare configuration, features a common proper motion white dwarf companion and is, therefore, a valuable target for understanding the formation and evolution of planetary systems.

\end{abstract}


\section{Introduction}
\label{sec:Introduction}

As of November 2021, there are more than 4800 known extrasolar planets (exoplanets), of which over 900 have been discovered with the radial velocity (RV) method utilizing the Doppler effect\footnote{up to date statistics available at \url{http://exoplanet.eu}}.
The Doppler spectroscopy remains the
best way for astronomers to determine the exoplanet orbital geometry, eccentricity, and planetary mass distribution, which are fundamentally important for understanding planet formation and evolution.
In radial velocity surveys for extrasolar planets, sparse phase coverage of the orbit may lead to model ambiguity,
such that two near-resonant companions can mask as a single one in an eccentric orbit \citep[see][]{Escude2010,Wittenmyer2013,Kuerster2015,Boisvert2018}. With steadily improving instruments, data analysis methods, and more precise data, it is worthwhile revisiting targets with already known planetary systems, as these could reveal additional undetected planets \citep[see, e.g.][]{Marcy2001,Trifonov2017,Wittenmyer2019}.

We initiated an RV follow-up survey of single exoplanet systems discovered with the Doppler method, targeting systems announced to have exoplanets in moderately eccentric orbits based on relatively sparse RV datasets, for which an alternative multi-planet model cannot be excluded \citep{Kuerster2015}. Our survey aims to obtain more precise RVs at critical orbital phases and re-analyze the orbital configuration looking for evidence of additional planets.
Of particular interest to us are multi-planetary systems in low-order mean motion resonance (MMR), such as GJ\,876 \citep{Marcy2001,Rivera2010,Trifonov2018a}, HD\,82943 \citep{Tan2013}, $\eta$ Ceti \citep{Trifonov2014}, and other massive pairs \citep[see, e.g.,][]{Trifonov2019b}. The strongly resonant architecture of such multi-planetary systems speaks in favor of planet migration, which makes
their dynamical characteristics important elements for a comprehensive understanding of planet formation.

We present a new orbital analysis of the known planetary system around the G-dwarf star HD\,107148, using archival Keck HIRES \citep{Vogt1994} and newly obtained high-precision HARPS \citep{Mayor2003} data.
We find that our orbital period for HD\,107148\,b of  $P_b \sim77$\,d differs significantly from the literature value of $\sim48$\,d, which we attribute to an alias induced from the HIRES observational schedule.
We further find significant evidence for the existence of a second warm Neptune mass planet on an eccentric orbit ($e_c$ = 0.41) with an orbital period of $\sim18$\,d. We present a detailed orbital update of the now confirmed two-planet system.

The paper is organized as follows;
In \autoref{sec:Literature}, we present an overview of the literature of the system. In \autoref{sec:Data}, we present the data we use for our analysis. \autoref{sec:Analysis} describes our analysis methods and numerical setups and presents our main results. Summary and conclusions of our results are given in \autoref{sec:ConclusionDiscussion}. \\

\section{The HD 107148 system}
\label{sec:Literature}

\begin{table}[htp!]

\caption{Stellar parameters of HD\,107148 and, if available, their 1-$\sigma$ uncertainties.}
\label{table:phys_param}

\centering
\begin{tabular}{ l l r r}     
\hline\hline  \noalign{\vskip 0.5mm}
  Parameter   &HD\,107148   \hspace{20 mm} &  reference \\
\hline    \noalign{\vskip 0.5mm}
   Spectral type                            & G1V          & [1] \\
   $V$ [mag]                           &     8.01    & [2] \\
   $B-V$ [mag]                           & 0.71        & [2] \\
     Distance  [pc]                           & 49.486$_{-0.115}^{+0.115}$   & [3] \\   \noalign{\vskip 0.9mm}
                               & 49.416$_{-0.115}^{+0.116}$   & [4] \\   \noalign{\vskip 0.9mm}
   Luminosity    [$L{_\odot}$]              & 1.321     & [4] \\
   $T_{\mathrm{eff}}$~[K]                   & 5833 $\pm$ 34     & [5] \\
   $\log g~[\mathrm{\mathrm{cm\,s}}^{-2}]$       & 4.42 $\pm$ 0.24    & [5] \\
   {}[Fe/H]                                 & 0.34 $\pm$ 0.07   & [5]  \\
   Mass    [$M_{\odot}$]                    & 1.127$_{-0.025}^{+0.035}$    & [5]\\ \noalign{\vskip 0.9mm}
   Radius    [$R_{\odot}$]                  & 1.15$_{-0.05}^{+0.05}$    & [5]  \\ \noalign{\vskip 0.9mm}
   Age    $[$Gyr$]$                         & 3.5$_{-1.3}^{+1.5}$     &  [5]\\

   $v\sin{i}$    [${\mathrm{km\,s}}^{-1}$]              & 1.36$\pm0.210$     & [5] \\
   Parallax [mas]                   & 20.25 $\pm$ 0.03     & [6] \\

\hline\hline \noalign{\vskip 0.5mm}

\end{tabular}

\tablecomments{\small 1 -- determined according to \citet{Pecaut2013},  2 -- \citet{ESA1997}, 3 -- \citet{Gaia2018b}, 4 -- \citet{Bailer-Jones2018}, 5 -- \citet{SotoJenkins2018}, 6 -- \citet{Gaia2021}}

\end{table}

\subsection{Stellar parameters}
\label{sec:Literature_1}

Literature stellar parameters and their 1-$\sigma$ uncertainties for HD\,107148 (HIP\,60081, BD-02 3497, TIC\,66666079), are listed in \autoref{table:phys_param}. Briefly, HD\,107148 is a Solar-type star
with a stellar mass $M_\star$ = $1.127^{+0.035}_{-0.025}$\,$M_{\odot}$ and stellar radius $R_\star$ = 1.15 $\pm$ 0.05\,$R_{\odot}$ \citep{SotoJenkins2018}. In addition to age, mass, radius, and effective temperature, \citet{SotoJenkins2018} determined that HD\,107148 is a super metal-rich star with [Fe/H] = 0.34 $\pm$ 0.07. Precise parallax  data obtained from the {\em Gaia} mission \citep{Gaia2016,Gaia2018b}, indicate that HD\,107148 is relatively close to the Sun with an estimated distance of $d$ = 49.486 $\pm$ 0.115\,pc.

 \citet{Tokovinin2012} reported the possible detection of a white dwarf (WD) stellar companion with a projected separation of about 1790\,au from the main sequence star HD\,107148. Later, \citet{Mugrauer2014} confirmed the WD companion as HD\,107148\,B (we continue to call the main sequence star HD\,107148 without adding an extra "A"). For the WD companion \citet{Mugrauer2016} obtained a mass of $M_{WD}$ = 0.56 $\pm$ 0.05$\,M_{\odot}$, an effective temperature of $T_{\rm eff}$ = 6150 $\pm$ 250\,$K$, a cooling age of 2.1 $\pm$ 0.27\,Gyr, a surface gravity of $\log{g}=7.95\pm 0.09\,\mathrm{cm\,s^{-2}}$ and a luminosity of $L=(2.0\pm 0.2)\times10^{-4}L{_\odot}$. By assuming the initial-to-final-mass-relation for white dwarfs from \cite{Catalan2008}, \cite{Mugrauer2016} derived a progenitor mass of $1.4\pm0.6\mathrm{M_{\odot}}$, leading to an estimated total age of $6.0\pm4.8$ Gyr for the WD companion.

 \subsection{Literature overview of the exoplanet system}
\label{sec:Literature_2}

Based on  35 Keck HIRES observations, obtained over a temporal baseline of six years (January 2000 to January 2006), \citet{Butler2006b} reported the discovery of
an exoplanet companion, HD\,107148\,b, with an orbital period of $P$ = 48.056 $\pm$ 0.057\,d, eccentricity of $e$ = 0.05 $\pm$ 0.17 and a mass of $m\sin{i}$ = 0.210 $\pm$ 0.036 $M_{\rm jup}$. Later, with more HIRES data available (a total 60 observations from January 2000 to January 2014), \citet{Butler2017} reported a significant periodicity in HD\,107148\,b of $P$ = 77.26 $\pm$ 0.09\,d\footnote{This value does not appear in recent databases, but appears in the supplementary Table 2 of \citet{Butler2017}}, which does not match their earlier result of 48\,d. In this work, we show that the signal of $\sim48$\,d is an alias of the true planetary period of HD\,107148 b, which is  successfully recovered by \citet{Butler2017} and our work, using more HIRES and HARPS data (see \autoref{sec:Periodogram}).

HD\,107148 has been observed as part of campaign 10 of the Kepler K2 mission \citep{Howell2014}. Our inspection of the light curves did not reveal any planetary transit signal.
During the revision of this paper, we become aware that HD\,107148 was observed by the Transiting Exoplanet Survey Satellite \citep[TESS,][]{Ricker2015} in sector 46. We inspected the light curve and found no transit signal in the TESS photometry window. Further, we found the TESS light curve stable over sector 46, suggesting that HD\,107148 is likely a photometrically quiet star.

During the advanced stage of this work, we became aware of the results from the \textit{The California Legacy Survey} \citep{Rosenthal2021,Fulton2021}, which includes HD\,107148. In particular, \citet{Rosenthal2021} present an updated orbital catalog of 178 planets, based on a large collection of RVs obtained with the Keck-HIRES \citep{Vogt1994}, APF-Levy \citep{Vogt2014}, and Lick-Hamilton \citep{Fischer2013} spectrographs over the past three decades. In their paper, they correctly identified the frequency of $\sim48$\,d as an alias of the actual period of HD\,107148 b ($\sim77$\,d), and also reported evidence of a new Neptune-mass planet
with a period of $P_{\rm c}$ $\sim18.33$\,d, which is in excellent agreement with our conclusions reported in this work. Our work differs from that of \citet{Rosenthal2021} in that we also perform a detailed orbital, activity, and dynamical analysis of the system. Also the orbital update reported in  \citet{Rosenthal2021} is based on old and newly obtained proprietary HIRES RVs, whereas our independent discovery of HD\,107148 c is based on newly obtained HARPS data in addition to the literature HIRES data set from \citet{Butler2017}. We consider the independent work of both teams and their mutually agreeing results to be solid evidence of the existence of the  HD\,107148 b \& c planets.

\section{Data}
\label{sec:Data}

\subsection{HIRES data}
\label{sec:HIRES}

We found 60 precise RV measurements of HD\,107148 obtained with the HIgh Resolution Échelle Spectrograph, mounted to the 10-m telescope of Keck Observatory, Hawaii, USA, \citep[HIRES,][]{Vogt1994}. HIRES is a general-purpose spectrograph, which relies on  the iodine (I$_{\rm 2}$) cell technique \citep{Marcy1992} to achieve radial velocity measurements with a precision of about 3\,m\,s$^{-1}$ \citep{Butler1996}.

The HIRES RV data of HD\,107148 were made publicly available in the \citet{Butler2017} catalog of precise Doppler measurements and stellar-line activity-index measurements, which consists of $\sim 65\,000$ HIRES spectra for $\sim1700$ stars obtained between 1996 and 2014.
Shortly after \citet{Butler2017} published their RV collection, \citet{TalOr2018} re-analyzed and corrected the HIRES data set for small yet significant systematic nightly zero-point variations in the data, increasing the precision. The magnitude of these corrections is of the order of $\sim1$\,m\,s$^{-1}$ and does not strongly affect our orbital analysis, yet we decided to use the more precise RV data set released by \citet{TalOr2018}.
From these data, we omitted one outlier taken at epoch BJD = 2451704.828, by performing a 3-$\sigma$-clipping to the HIRES data set.\looseness=-9

The HIRES data were obtained between January 2000 and January 2014 with a total temporal baseline of 5122\,d and have a weighted root mean square of wrms$_{\rm HIRES}$ = 6.7\,m\,s$^{-1}$ and a median RV uncertainty of $\hat\sigma_{\rm HIRES}$ = 1.5\,m\,s$^{-1}$, respectively.

 \subsection{HARPS data}
 \label{sec:HARPS}

We obtained a total of 92 high signal-to-noise spectra of HD\,107148 with the High Accuracy Radial velocity Planet Searcher spectrograph \citep[HARPS][]{Mayor2003}, mounted at the ESO 3.6\,m Telescope, La Silla, Chile.
We also found a total of 13 archival HARPS spectra of HD\,107148 in the official ESO archive\footnote{\url{http://archive.eso.org/wdb/wdb/adp/phase3_main/form}}.
Thus, for HD\,107148, we have a total of 105 HARPS spectra from which we extracted high-precision RVs for our orbital analysis.

We derived precise RV measurements and spectral activity indices with the SpEctrum Radial Velocity AnaLyser \citep[SERVAL,][]{Zechmeister2018} pipeline, as well as with the
official HARPS-DRS pipeline.
From SERVAL we obtained the time series of the stellar activity indicators chromatic index (CRX), differential line width (dLW), H$\alpha$, and Na~D$_1$ and~D$_2$ \citep[see][for more details]{Zechmeister2018},
whereas from the DRS, we obtained the cross-correlation function (CCF), full width half maximum FWHM, the CCF Bisector slope (BIS), and contrast (CON), which are important stellar activity indices \citep[e.g.][]{Queloz2001}.
We note that in May 2015, the HARPS spectrograph underwent a major upgrade of the fiber link between the
telescope focal plane and the spectrograph entrance \citep[see,][]{LoCurto2015}.
This resulted in a significant change of the instrumental profile, which leads to a notable offset between the pre- and post-upgrade RVs and activity indices.
Thus, following \citet{Trifonov2020} we ran SERVAL independently on the post and pre-fiber upgrade data, and we always treated
the pre- and post- HARPS-DRS and SERVAL results as taken from two independent instruments.

The HARPS data were obtained between February 2005 and July 2019 with a total temporal baseline of 5257\,d.
The pre fiber upgrade RVs have a weighted root mean square of wrms$_{\rm HARPS-pre}$ = 8.34\,m\,s$^{-1}$ and a median RV uncertainty of $\hat\sigma_{\rm HARPS-pre}$ = 1.04\,m\,s$^{-1}$,
whereas the post upgrade data have wrms$_{\rm HARPS-post}$ = 7.83\,m\,s$^{-1}$ and $\hat\sigma_{\rm HARPS-post}$ = 0.65\,m\,s$^{-1}$, respectively.

Since most of our spectra were taken in three consecutive measurements per visit, we decided to average out the RV measurements in nightly-averaged bins in order to compensate for the contribution of short-term, stellar p-mode pulsations \citep[e.g.][]{Dumusque2011c, Chaplin2019}. As a result, for our orbital analysis, we used nine nightly averaged RV measurements taken prior to the fiber upgrade and 32 nightly averaged measurements taken after the fiber upgrade.\\
The nightly-averaged RVs of the pre-upgrade HARPS data have a weighted root mean square (wrms) residual of wrms$_{\rm HARPS-pre}$ = 8.3\,m\,s$^{-1}$ and a median RV uncertainty of $\hat\sigma_{\rm HARPS-pre}$ = 1.0\,m\,s$^{-1}$,
whereas the post-upgrade HARPS data have wrms$_{\rm HARPS-post}$ = 7.74\,m\,s$^{-1}$ and $\hat\sigma_{\rm HARPS-post}$ = 0.66\,m\,s$^{-1}$, respectively.
The HARPS-SERVAL Doppler measurements, activity index data, and their individual formal uncertainties are available in \autoref{tab:HARPS_1} and \autoref{tab:HARPS_2}.

   \begin{figure}
    \centering
    \includegraphics[width=9cm]{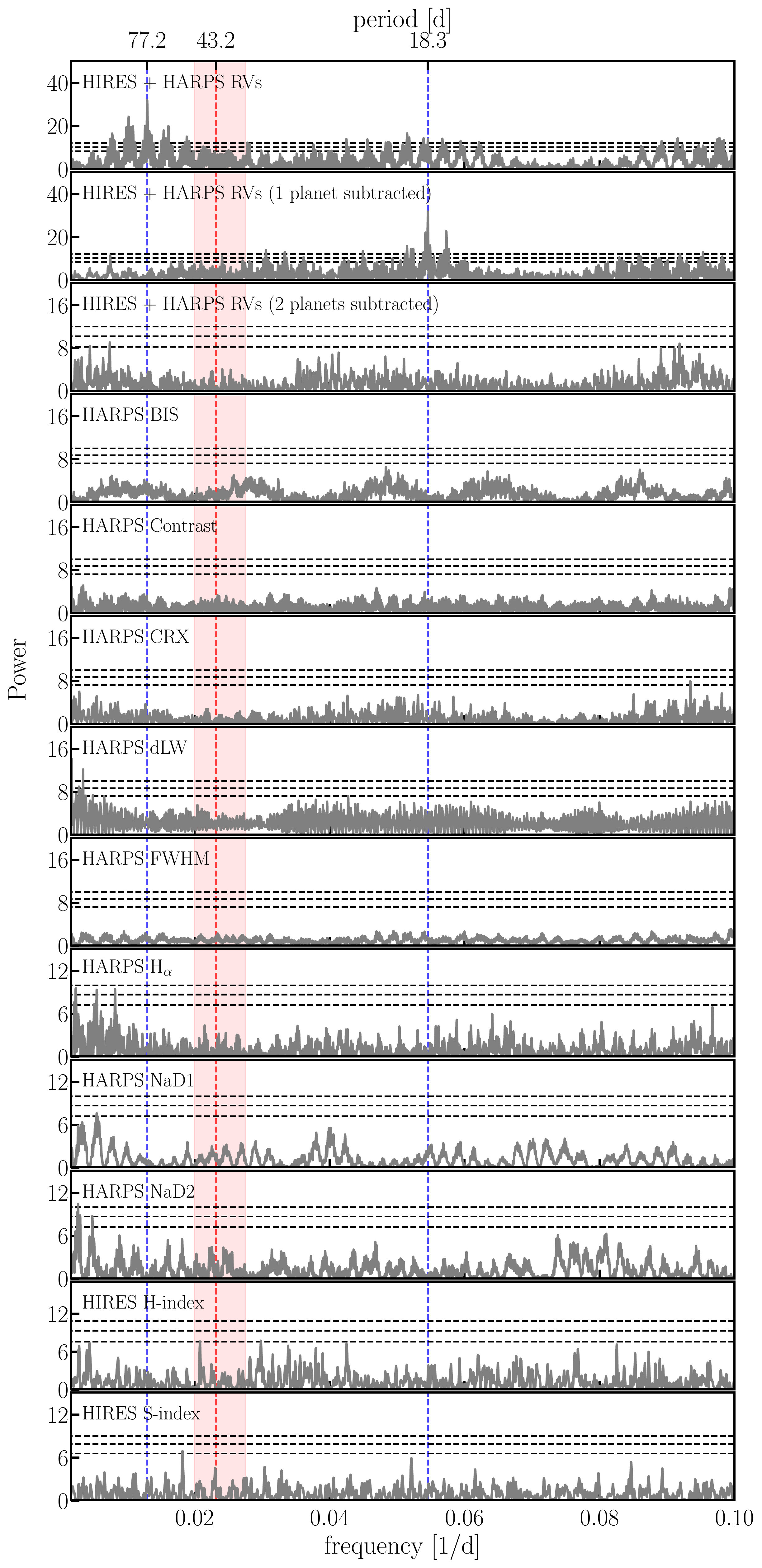}
     \caption{GLS power spectrum for RV data obtained from Keck HIRES and HARPS, as well as stellar activity data obtained from HIRES and HARPS. Horizontal dashed lines indicate FAP levels of 10\%, 1\%, and 0.1\%. Blue vertical dashed lines show the periods of the two significant signals. An upper limit for the most likely rotational period of HD\,107148 is indicated with a red vertical dashed line, whereas the 1-$\sigma$ uncertainty frequencies of the stellar rotation are shown with a red-transparent region.}
    \label{fig:activity_periodograms}
\end{figure}

\subsection{CARMENES data}
\label{Sec3.2}

We obtained 12 high signal-to-noise spectra of HD\,107148 with the CARMENES\footnote{CARMENES stands for; Calar Alto high-Resolution search for
M dwarfs with Exo-earths with Near-infrared and optical Echelle Spectrographs. See, \url{https://carmenes.caha.es/}} instrument \citep[][]{Quirrenbach2016,Reiners2018a},
mounted on the 3.5\,m telescope in Calar Alto, Spain.
These spectra were acquired on 15 January 2017, 1 March 2017, and 11 April 2017.
We derived precise CARMENES optical and near-IR RV measurements and activity indicators using the SERVAL pipeline, in the same way we did for HARPS (see \autoref{sec:HARPS}).

The nightly averaged data resulted in only three precise CARMENES RV measurements in the optical and three in the near-IR, respectively. We find that the small number of CARMENES RVs is insufficient for a significant contribution to our orbital solution, thus, these were not used in combination with HARPS and HIRES. Nonetheless, we investigated the consistency of the CARMENES data with respect to the orbital solution derived with the HIRES and the HARPS RV data sets (see \autoref{sec:OrbDynAnalysis}).
The individual CARMENES Doppler measurements, activity indices, and their individual formal uncertainties are available in  \autoref{tab:CARM_VIS} and \autoref{tab:CARM_NIR}.\looseness=-8

\section{Analysis and Results}
\label{sec:Analysis}
 \subsection{Periodogram analysis}
 \label{sec:Periodogram}

 \begin{figure}
    \centering
    \includegraphics[width=0.232\textwidth]{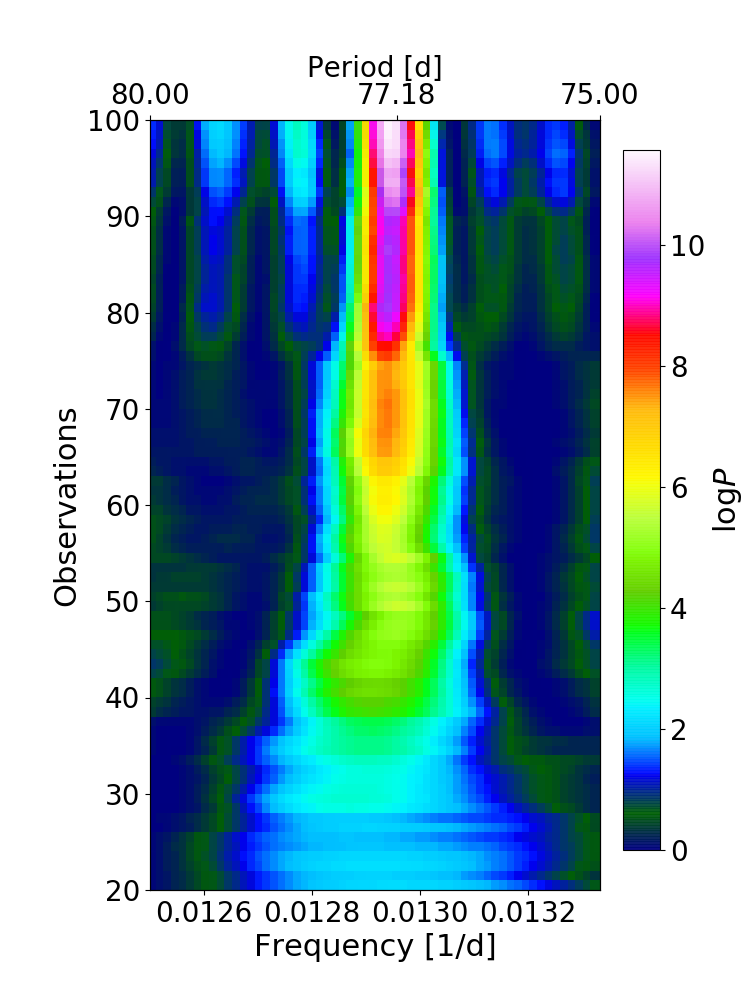} \includegraphics[width=0.232\textwidth]{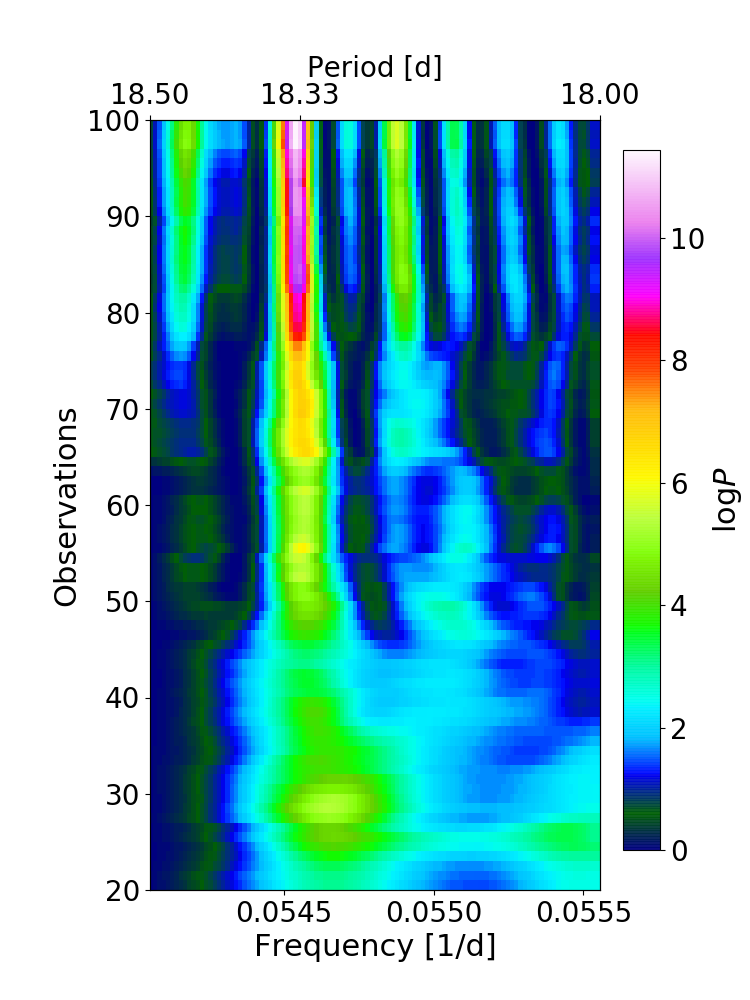}
    \caption{Stacked-Bayesian GLS periodogram of the RV data of HD\,107148 focused at the 77\,d signal (left), and on the signal at 18\,d (right). The s-BGLS of the 77\,d signal is calculated by fitting for the RV offset between HIRES, HARPS-pre, and HARPS-post. Similarly, the s-BGLS of the 18\,d signal was calculated on the RV residuals of a fit where the signal at 77\,d has been removed.}
    \label{fig:sBGLS}
\end{figure}

We inspected the available RV and activity indicator data for significant periodicity using the generalized Lomb-Scargle periodogram \citep[GLS;][]{Zechmeister2009}. As our significance threshold, we adopted the analytically calculated false-alarm probability (FAP) level of $<$0.1\% as defined in \citet[][]{Zechmeister2009}.
\autoref{fig:activity_periodograms} shows the resulting GLS periodograms for the radial velocities and the stellar activity time series.
The ordinate axis indicates the ZK GLS power \citep[][]{Zechmeister2009}, the abscissa indicates the scanned frequency, while horizontal dashed lines indicate FAP levels of 10\%, 1\%, and 0.1\% from bottom to top.
The first three panels show the GLS periodograms of the combined HARPS + HIRES  RVs after their mutual RV offset and significant signals were subtracted.  This being said, the first three panels represent residual GLS periodograms of; a flat-model (i.e., only RV offsets modeled), one-planet fit, and a two-planet fit, respectively. The two top GLS periodograms show two significant signals, indicated in all panels with vertical blue-dashed lines. The flat-model residuals of the HIRES and HARPS data indicate a significant period of $77.2$\,d, which is consistent with that reported in \cite{Butler2017}. We are, however, unable to recover the $48$\,d period originally reported by \cite{Butler2006b}. We conclude that the literature period of $P_b \sim48$\,d for HD\,107148 b is an aliasing artifact of the new (true) $77.2$\,d planetary signal and the $\sim29$\,d period of the lunar cycle, which must have affected the HIRES observation schedule. Evidence for that is the strong 29.6\,d peak in the HIRES window function shown in \autoref{Fig:HIRESwindow}. Indeed, P$_{\rm alias}$ = $1/ (f_{\rm WF}$ -- $f_{\rm pl.}) = 1/ (\frac{1}{29.6\,\rm d} - \frac{1}{77.2\,\rm d})$ = 48\,d.\looseness=-9

After a Keplerian model fits the 77.2\,d signal, its residuals show a second significant period around $\sim18.3$\,d, which we interpret as evidence of a possible second planetary companion in the HD\,107148 system. We find that the 18.3\,d signal becomes evident thanks to the additional HARPS RVs. After performing a simultaneous two-planet Keplerian fit using these periods as an initial guess, no significant power remains in the residuals.

In \autoref{fig:activity_periodograms} the most probable stellar rotational period and its 1-$\sigma$ range of frequencies are indicated with a red vertical dashed line and red-region. These were calculated as $P_{\rm rot}/ \sin{i} = 2\pi R_{\star} / v\sin{i}$.
Using the radius and projected rotational velocity from \autoref{table:phys_param}, we derived an upper limit for the stellar rotation period $P_{\rm rot}=43.2\pm6.9$ d, which as \autoref{fig:activity_periodograms} indicates is sufficiently far in frequency space from any of the planetary signals.
Nevertheless, the estimated rotational period of the star only represents an upper limit for the actual rotational period due to the unknown inclination of the system. At this point of the GLS analysis, we assumed that there is still a possibility that the significant $\sim18$ d signal could be induced by stellar activity.
Therefore, as a mandatory sanity check, we inspected the HIRES and HARPS stellar activity indicators to check for possible effects caused by the star itself that could explain the 18\,d and the 77\,d signals.

         \begin{table}[ht]
    
    \centering    
    \caption{{Best-fit parameters obtained from a Simplex MLE optimization and errors determined from an MCMC analysis, assuming a coplanar egde-on configuration.}}   
    \label{table:MCMC_results}      
    
    \begin{tabular}{lrrrrrrrr}     
    
    \hline\hline  \noalign{\vskip 0.7mm}      
    Parameter \hspace{0.0 mm}& Planet c & Planet b \\
    \hline \noalign{\vskip 0.7mm} 

        $K$ [m\,s$^{-1}$]             &       5.28$_{-0.45}^{+0.32}$ &       8.75$_{-0.41}^{+0.38}$ \\ \noalign{\vskip 0.9mm}
        $P$ [day]                     &      18.3270$_{-0.0016}^{+0.0008}$ &      77.185$_{-0.025}^{+0.01}$ \\ \noalign{\vskip 0.9mm}
        $e$                           &       0.40$_{-0.08}^{+0.04}$ &       0.15$_{-0.06}^{+0.02}$ \\ \noalign{\vskip 0.9mm}
        $\omega$ [deg]                &     308$_{-10}^{+14}$ &     223$_{-16}^{+19}$ \\ \noalign{\vskip 0.9mm}
        $M_{\rm 0}$ [deg]             &      60$_{-16}^{+5}$ &     318$_{-24}^{+14}$ \\ \noalign{\vskip 0.9mm}
        $a$ [au]                      &       0.1415$_{-0.0015}^{+0.0015}$ &       0.3692$_{-0.0038}^{+0.0037}$ \\ \noalign{\vskip 0.9mm}
        $m \sin i$ [$M_{\rm jup}$]    &       0.068$_{-0.005}^{+0.004}$ &       0.196$_{-0.009}^{+0.009}$ \\ \noalign{\vskip 0.9mm}
        RV$_{\mathrm{off,\,HIRES}}$ [m\,s$^{-1}$]&       \multicolumn{2}{c}{1.86$_{-0.37}^{+0.63}$} \\ \noalign{\vskip 0.9mm}
        RV$_{\mathrm{off,\,HARPS (pre)}}$ [m\,s$^{-1}$]&       \multicolumn{2}{c}{-9.97$_{-0.65}^{+0.36}$} \\ \noalign{\vskip 0.9mm}
        RV$_{\mathrm{off,\,HARPS (post)}}$ [m\,s$^{-1}$]&      \multicolumn{2}{c}{0.89$_{-0.37}^{+0.28}$} \\ \noalign{\vskip 0.9mm}
        RV$_{\mathrm{jit,\,HIRES}}$ [m\,s$^{-1}$]&       \multicolumn{2}{c}{3.28$_{-0.27}^{+0.54}$} \\ \noalign{\vskip 0.9mm}
        RV$_{\mathrm{jit,\,HARPS (pre)}}$ [m\,s$^{-1}$]&       \multicolumn{2}{c}{0.00$_{-\,-0.31}^{+1.78}$} \\ \noalign{\vskip 0.9mm}
        RV$_{\mathrm{jit,\,HARPS (post)}}$ [m\,s$^{-1}$]&       \multicolumn{2}{c}{1.17$_{-\,-0.01}^{+0.53}$} \\ \noalign{\vskip 0.9mm}
        $\chi^2$                      &      \multicolumn{2}{c}{98.67} \\
        $\chi_{\nu}^2$                &       \multicolumn{2}{c}{1.17} \\
        $rms$ [m\,s$^{-1}$]           &       \multicolumn{2}{c}{2.91} \\
        $wrms$ [m\,s$^{-1}$]          &       \multicolumn{2}{c}{2.45} \\
        $-\ln\mathcal{L}$             &    \multicolumn{2}{c}{-224.98} \\
        N$_{\rm RV}$ data             &        \multicolumn{2}{c}{100} \\
        Epoch                         & \multicolumn{2}{c}{2451553.08} \\
        \\
    \hline \noalign{\vskip 0.7mm}

    \end{tabular}



    \end{table}

The remaining periodograms in \autoref{fig:activity_periodograms} show the HIRES and HARPS stellar activity indicators as labeled in the panels. There are no significant periods in the stellar activity data that coincide with the planetary orbital frequencies. However, it is worth noting that there are peaks in the HARPS' dLW, H$\alpha$, and NaD2 that can be considered significant at low frequencies, which could mean that HD\,107148 is a somewhat weakly active star.

 As an additional test to our GLS analysis of the available activity indicators, we investigated the coherence of the two significant RV signals around 18\,d and 77\,d, by performing a stacked-Bayesian GLS periodogram test \citep[s-BGLS;][]{Mortier2015, Mortier2017} of the RV data.
 \autoref{fig:sBGLS} shows the results from the s-BGLS analysis. To construct the s-BGLS periodograms, we applied the normalization as in \cite{Mortier2017} and fitted for the mutual RV offsets between the HIRES, the HARPS-pre, and the HARPS-post data. Both signals show growth of their signals probability over the complete time-span of the observations, which is a strong indication for their planetary origin. Therefore, we conclude that the 77\,d and 18\,d signals are of planetary nature, and we further base our Keplerian analysis on adopting the two-planet hypothesis for HD\,107148.

\begin{figure*}
    \centering
    \includegraphics[width=17.5cm]{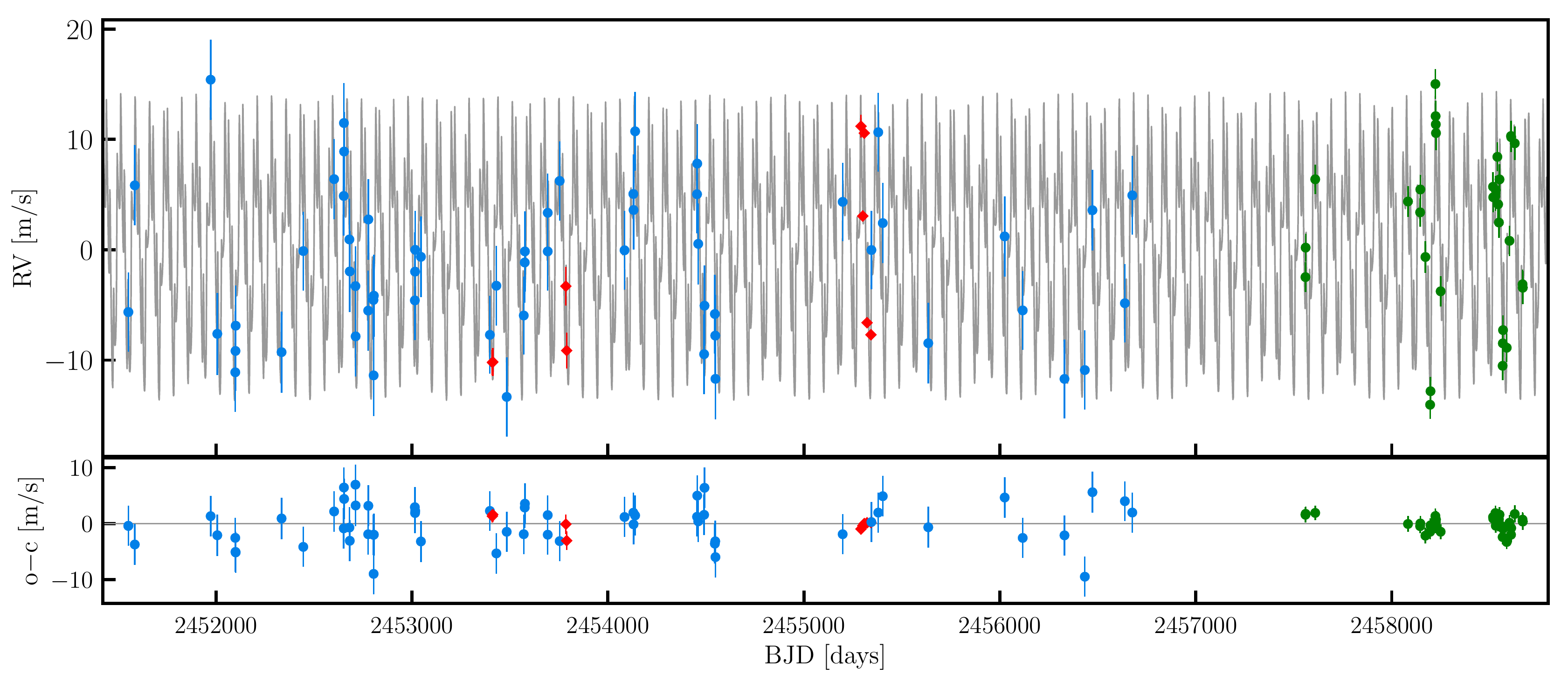} \\
    \includegraphics[width=17.5cm]{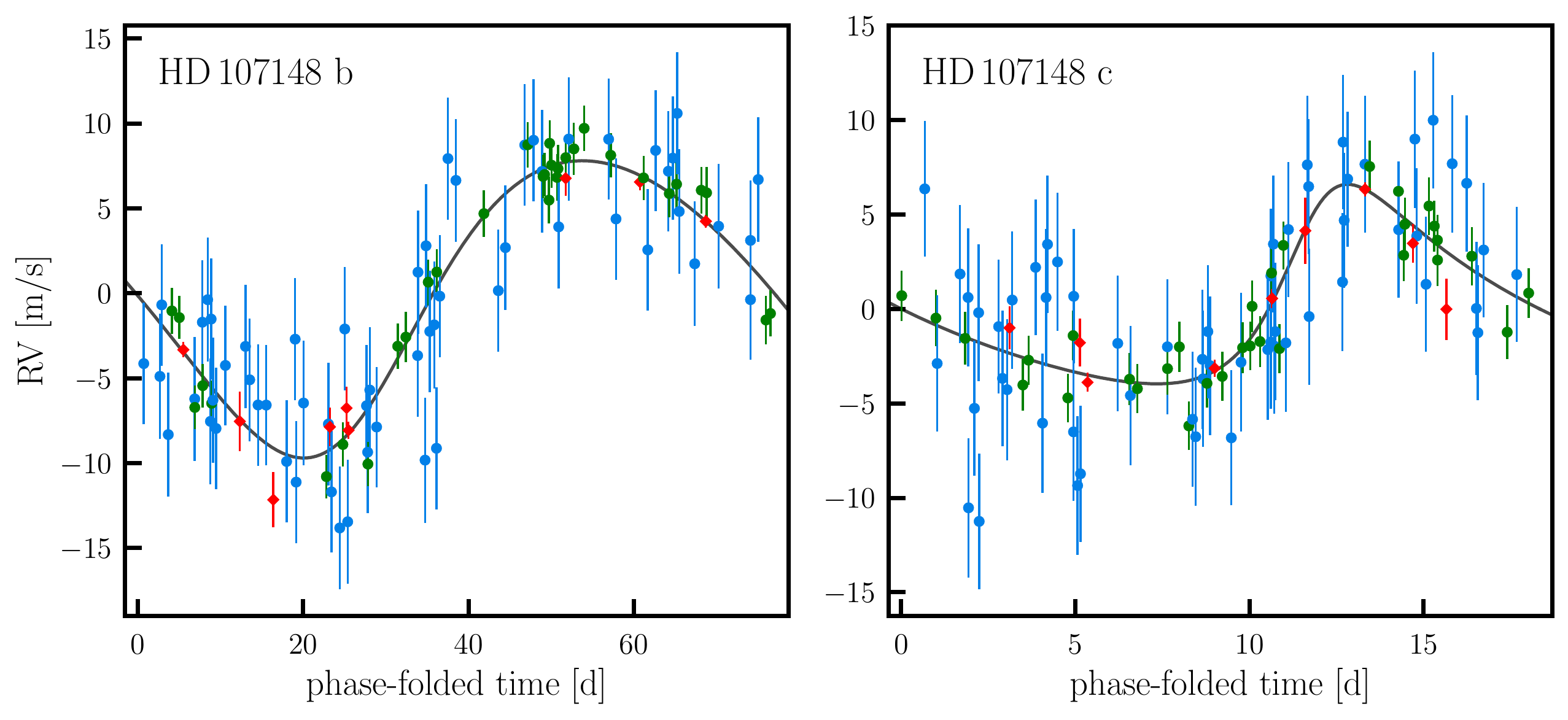}
    \put(-200,150){  }

    \caption{
    {\em Top panel}: Doppler radial velocity measurements of HD\,107148 from HIRES (blue) and HARPS (red and green, pre and post fiber correction, respectively) modeled with a two-planet Keplerian model (black curve) along with the corresponding residuals beneath.
    {\em Bottom panels}: Phase-folded planetary signals fitted to the data, after the respective other planetary signal is subtracted. Left: HD\,107148\,b, right: HD\,107148\,c. The data uncertainties include the estimated RV jitter.
}
\label{fig:kepler_fits}
\end{figure*}

 \begin{figure}
    \centering
    \includegraphics[width=8.5cm]{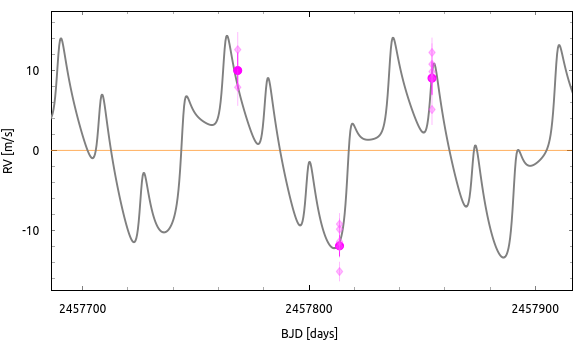}
     \caption{Radial velocity measurements of HD\,107148 from CARMENES fitted  with a variable RV offset to the best-fit, two-planet Keplerian model constructed from the HARPS and HIRES RV data (see \autoref{fig:kepler_fits}).
     The individual consecutive measurements are shown with transparent magenta diamonds, whereas the nightly averaged RV measurements are shown with filled magenta circles.}
    \label{fig:CARM}
\end{figure}

\subsection{Orbital analysis}
\label{sec:OrbDynAnalysis}

To model the planetary orbits from the available precise Doppler data of HD\,107148, we used the
\texttt{Exo-Striker}\footnote{\url{https://github.com/3fon3fonov/exostriker} } \citep{Trifonov2019_es} exoplanet toolbox.
  We used a  Nelder-Mead (Simplex) algorithm \citep[][]{NelderMead}, which  optimizes the likelihood ($-\ln{\mathcal{L}}$) value of
  the adopted RV model. This is a classical maximum Likelihood Estimator (MLE) scheme, which converges to the maximum $-\ln{\mathcal{L}}$ value, returning the "best-fit" Keplerian orbital parameters. The \texttt{Exo-Striker} offers two Keplerian modeling schemes; a pure unperturbed multi-Keplerian model, and a more complex N-body algorithm, which models the RVs by calculating the mutual gravitational perturbations between the massive bodies in the system \citep[see, e.g.,][]{Tan2013,Trifonov2014,Trifonov2019b}. The latter model seems appropriate in the context of the warm-pair of relatively massive planets evident in the HD\,107148 data, where dynamics cannot be neglected. However, for accurate orbit calculation, the N-body model requires a sufficiently small integration time-step of at least P$_{\rm inner~planet}$/100, or $dt_{\rm N-body} \sim0.18$ \,d, which, over the temporal baseline of the RV data of $\sim20$ \,yr, makes this model computationally too expensive.
 It is worth noting that we performed a quality test of a two-planet Keplerian MLE and alternative N-body MLE best-fits. We did not find a significant difference in their $\ln{\mathcal{L}}$, meaning that the RV data does not have the needed quality to constrain the dynamical properties of the system, or the secular dynamical time scales of the system are much longer than the temporal baseline of the HIRES and HARPS data (see details in \autoref{sec:DynamicalAnalysis}). Thus, we find that the pure multi-Keplerian model is entirely sufficient to precisely describe the orbital configuration of HD\,107148, given the available RV data, and it was adopted in this work.

 Nonetheless, for testing coplanar inclined orbital geometries, we naturally applied an N-body model MLE fitting. Mutually inclined configurations are very ambiguous and cannot be effectively constrained given the available RV data, and are thus not considered in our analysis.\looseness=-4

In our adopted MLE scheme, we fitted for the zero point RV offsets of the HIRES, the HARPS-pre, and the HARPS-post data, and the spectroscopic Keplerian parameters of each planet included in the modeling, namely the RV semi-amplitude $K$, the orbital period $P$, the eccentricity $e$, the argument of periastron $\omega$ and the mean anomaly $M_0$. Since we adopted a pure Keplerian two-planet model, the planetary inclinations $i$ and the difference of the orbital ascending node $\Delta \Omega$ cannot be accessed even if the RV data is sensitive to these parameters\footnote{In rare cases of strongly interacting planetary systems, an N-body model can constrain the orbital alignment. See, e.g., the case of the GJ\,876 system \citep{Rivera2010,Nelson2016}}.  Therefore, we assumed only a coplanar and edge-on configuration of the HD\,107148 system (i.e., $i$=90$^\circ$ and mutual orbital inclination $\Delta i$=0$^\circ$). Following \cite{Baluev2009}, we also fitted for the white-noise variance of each RV data set. We added the RV "jitter" term in quadrature to the nominal RV uncertainties, and it also presents a penalty-term in our model's likelihood \citep[see][for the functional form of the $\ln{\mathcal{L}}$]{Baluev2009}.

We evaluated the posterior probability distribution of the fitted parameters, by adopting the affine-invariant ensemble Markov chain Monte Carlo (MCMC) sampler \citep{Goodman2010} using the
\texttt{emcee} package \citep{Foreman-Mackey2013}, which is integrated within the \texttt{Exo-Striker} tool.
We explored the parameter space by adopting non-informative flat priors, the range of which we estimated experimentally during the course of our MLE minimization analysis. Since we were more interested in studying possible parameter correlations and estimating the parameter uncertainties from the posterior distribution, we ran the MCMC chains starting from the "best-fit" parameters returned by the MLE optimization scheme. In our MCMC scheme, we used a setup of 200 walkers, 1\,000 burn-in samples, which we discarded from the analysis, followed by 5\,000 MCMC samples from which we constructed the parameter posterior distribution.
 We evaluated the sampler acceptance fraction, which should be between 0.2 and 0.5,
to consider the MCMC chains converged \citep[see][]{Foreman-Mackey2013}.
 The 1-$\sigma$ parameter uncertainties in this work were chosen as the 68.3\% confidence levels of the posterior parameter distribution.

    We have also tested the sensitivity of our obtained results due to possible correlated RV noise by applying a quasi-periodic Gaussian processes kernel together with a Bayesian nested sampling following an approach as in \cite{Stock2020}. We found no significant differences to our obtained results based on our MLE+MCMC scheme that does not incorporate an additional correlated noise model in the form of a Gaussian process. Hence, the omission of a correlated noise model can be justified in the case of HD\,107148. In particular, its inclusion has also not been favored significantly by the Bayesian log-evidence.

\subsection{Model comparison results}
\label{sec:OrbitalUpdate}

   \begin{table*}[htp]

\caption{Comparison between the quality of the fits for different models. Circular orbit means $e=0$, $\omega$ undefined and eccentric orbit means that $e$ and $\omega$ were kept as fit parameters.} 
\label{table:stat_param}    

\centering          
\begin{tabular}{ l l r r r r}     
\hline\hline  \noalign{\vskip 0.5mm}        
  Number & Model   \hspace{95mm} &$\ln{\mathcal{L}}$   &  BIC & $\Delta\ln\mathcal{L}$ & $\Delta$BIC\\  
\hline    \noalign{\vskip 0.5mm}                   
  0 & No planet                            & -337.89          & 689.60 & \dots & \dots \\ 
  1 & Only planet b (circular orbit)                         &     -289.35      & 606.33 & 48.54 & 83.27 \\ 
  
  2 & Only planet b (eccentric orbit)                         &     -287.35      & 611.53 & 50.54 & 78.07 \\ 
  
  3 & Two planets (both circular orbits)                         &     -245.31      & 532.07 & 92.58 & 157.53 \\ 
  
  4 & Two planets (only planet b eccentric orbit)                         &     -243.14      & 536.94 & 94.75 & 152.66 \\ 
  
  5 & Two planets (only planet c eccentric orbit)                         &     -232.14      & 514.93 & 105.75 & 174.67 \\ 
   
  6 & Two planets (both eccentric orbits)                         &     -224.98      & 509.82 & 112.91 & 179.78 \\

\hline\hline \noalign{\vskip 0.5mm}   

\end{tabular}
\end{table*}

 As a means to firmly constrain the orbital configuration of the system, we prepared different competing MLE model schemes with different degrees of freedom (and number of planets), and inspected the quality of their "best-fits". We performed two quality tests: (i) we compared the difference of the MLE fit's  likelihood as $\Delta\ln{\mathcal{L}} = |\ln{\mathcal{L}}_{\rm complex~model}| - |\ln{\mathcal{L}}_{\rm simpler~model}|$), (ii) we calculated and compared their Bayesian information criterion (BIC) values.  The BIC is defined as $\mathrm{BIC} =k\ln(n)-2\ln({\mathcal {L}})$, with the number of free parameters $k$, the number of measurements $n$ and the model's likelihood $\mathcal{L}$. Widely accepted values for a strong statistic evidence when comparing two models are $\Delta\ln{\mathcal{L}}>7$~\footnote{A relative probability of $R=e^{\Delta\ln\mathcal{L}}$ corresponds to $\sim0.1\%$ FAP for $\Delta\ln\mathcal{L}=7$, which is necessary for a significant detection \citep[see][]{Anglada-Escude2016}} and $\Delta\mathrm{BIC}>10$ \citep{Kass1995}. Thus we adopted these thresholds to claim a significant model improvement.  The results of this evaluation can be found in \autoref{table:stat_param}.

 First, we constructed our null model, for which free parameters are the RV offsets and jitters and which contains no planets (model "0").
Then, we injected a combination of Keplerian signals given the information from our GLS test (\autoref{sec:Periodogram}).  Alternative models "1" and "2" were constructed assuming only one planet with a period of $\sim77$\,d (HD\,107148\,b), whose orbital shape was set to be either strictly circular or allowed to be eccentric. In the circular model "1", the planet's eccentricity was forced to $e=0$ and its argument of periastron $\omega$ was undefined, while for the eccentric model "2", both $e$ and $\omega$, were kept as free fit parameters. Analogously, the following models "3" to "6" were constructed as the possible combinations of two planets with periods of $\sim77$\,d (HD\,107148\,b) and $\sim18$\,d (HD\,107148\,c) whose orbits were either strictly circular or allowed to be eccentric.

A comparison between models "2" and "6" shows that assuming a second planet greatly improves the fit, with the likelihood improving by $\Delta\ln\mathcal{L}=62.37$ and the BIC improving by $\Delta \mathrm{BIC}=101.71$.
 While there is a significant improvement for model "6" against models "3" and "4" according to our adopted thresholds, there is only weak evidence for model "6" against model "5" with $\Delta\ln\mathcal{L}=7.16$ and $\Delta \mathrm{BIC}=5.11$. Unlike the likelihood, which shows a significant improvement, the BIC takes the numbers of parameters into account, thus providing a more meaningful way of comparing the models. However, despite $\Delta\mathrm{BIC}$ not being sufficiently large enough to claim a significant improvement, we chose model "6" for the final orbital solution of this work and our further analysis, as it still maximizes both $\Delta\ln\mathcal{L}$ and $\Delta\mathrm{BIC}$ with respect to our null model.\\

 \autoref{table:MCMC_results} lists the best-fit parameters for our adopted model "6" with error estimates obtained from an MCMC as described in \autoref{sec:OrbDynAnalysis}.
  The posterior results of the MCMC analysis for our adopted model are shown in \autoref{Fig:corner}. From these, we could derive the planetary
  periods $P_b$ = 77.185$_{-0.025}^{+0.01}$\,d, $P_c$ = 18.3270$_{-0.0016}^{+0.0008}$\, eccentricities $e_b$ = 0.40$_{-0.08}^{+0.04}$,
  $e_c$ = 0.15$_{-0.06}^{+0.02}$, and masses $m_b\,\sin{i}$ = 0.196$_{-0.009}^{+0.009}$  $M_{\rm jup}$, $m_c\,\sin{i}$ = 0.068$_{-0.005}^{+0.004}$ $M_{\rm jup}$.
 Model "6" is further illustrated in the top panel of \autoref{fig:kepler_fits}, where the RV time series of both HARPS and HIRES data are shown together with the RV residuals on the bottom. The bottom two panels of \autoref{fig:kepler_fits} show the individual planetary signals, where the RVs were phase-folded with the best-fit orbital periods (HD\,107148\,b on the left and HD\,107148\,c on the right) after subtracting the other planet's signal.

 \autoref{fig:CARM} shows our CARMENES visual channel data (VIS) over-plotted on the best-fit model "6" from the HARPS and HIRES data. As we mentioned in \autoref{Sec3.2}, the CARMENES data are very sparse and taken only at three effective epochs. We find that their effective weight to the orbital solution is low, and thus they were not included in the orbital analysis. Nonetheless, we show the consistency of the CARMENES-VIS data to model "6" by fitting an RV offset and jitter, while all other parameters in model "6" were kept fixed.  The CARMENES near-IR data we obtain are showing much lower precision and were thus not tested.
  From \autoref{fig:CARM} the high precision of the CARMENES data can be assessed, which is in excellent agreement with our adopted model constructed by fitting the HIRES and the HARPS RVs. The nominal 11 CARMENES-VIS channel RVs have a median uncertainty of 1.79 m\,s$^{-1}$, and a weighted rms around the best-fit of $wrms$ = 2.41 m\,s$^{-1}$, whereas the three nightly averaged epochs have an RV  weighted rms of $wrms$ = 1.07 m\,s$^{-1}$, which is of the same order of precision as HARPS, and better than HIRES \citep[see also][for similar conclusions]{Trifonov2018a,Kaminski2018}. We note, however, that these estimates in the context of HD\,107148, are coming from low-number statistics. While the CARMENES data were not used in our orbital analysis, these RVs further strengthen our two-planet hypothesis of the HD\,107148 system due to their strong agreement with our two-planet, best-fit model and its uncertainties.

\begin{figure}
    \centering
    \includegraphics[width=8.8cm]{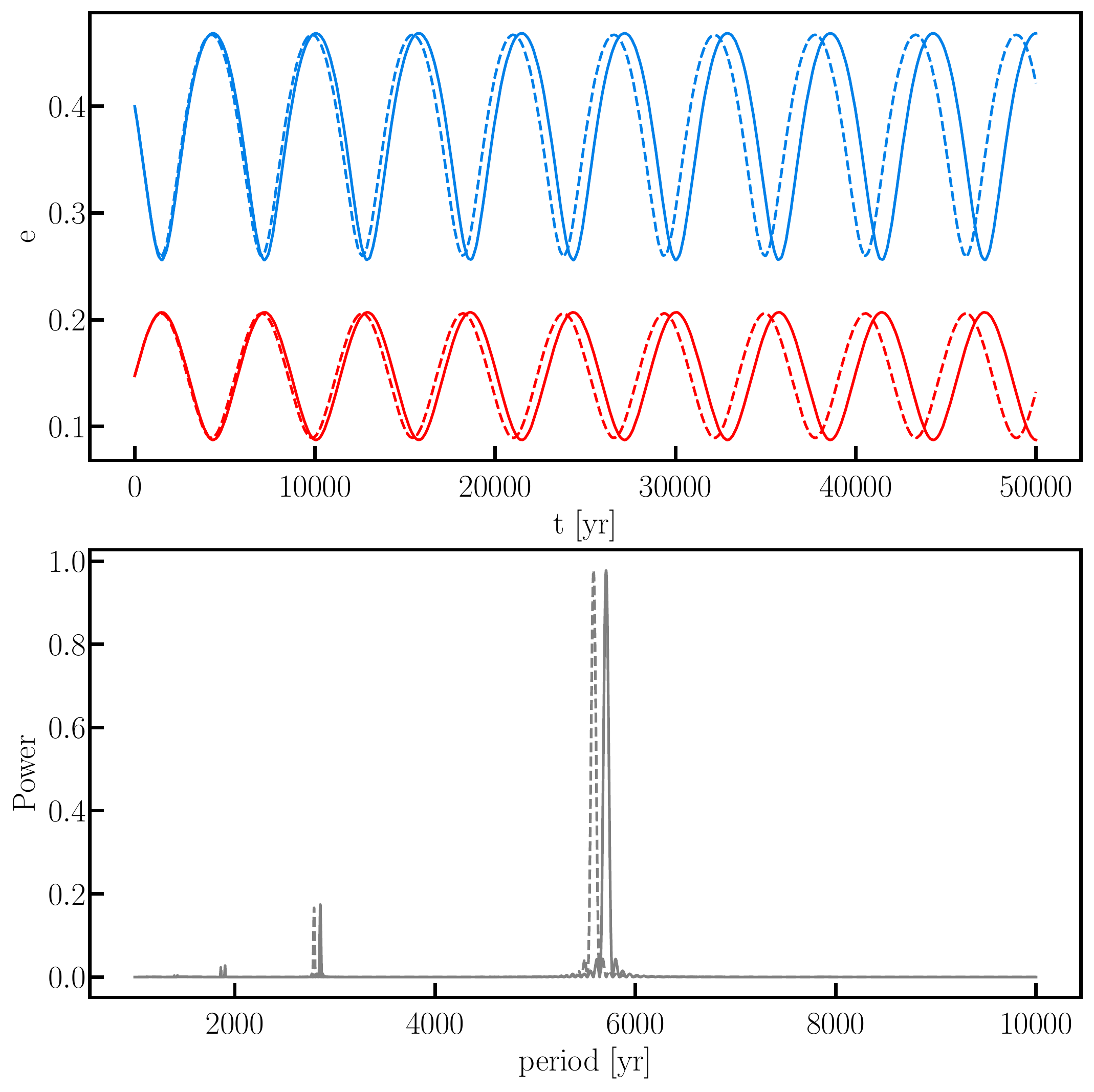} \\

    \caption{Eccentricity evolution of HD\,107148\,b (red) and HD\,107148\,c (blue) using an MVS algorithm with (dashed lines) and without (solid lines) general relativistic corrections.
    {\em Top panel}: Evolution of the eccentricities.
    {\em Bottom panel}: GLS periodograms of the eccentricity values, peaking at P = 5707.52 yr (MVS, solid line) and P = 5581.36 yr (MVS-GR, dashed line). The smaller, shorter period GLS peaks are high-frequency harmonics of the true periods. }
    \label{fig:orbital_evolution}
\end{figure}

\begin{figure*}
    \includegraphics[width=12cm]{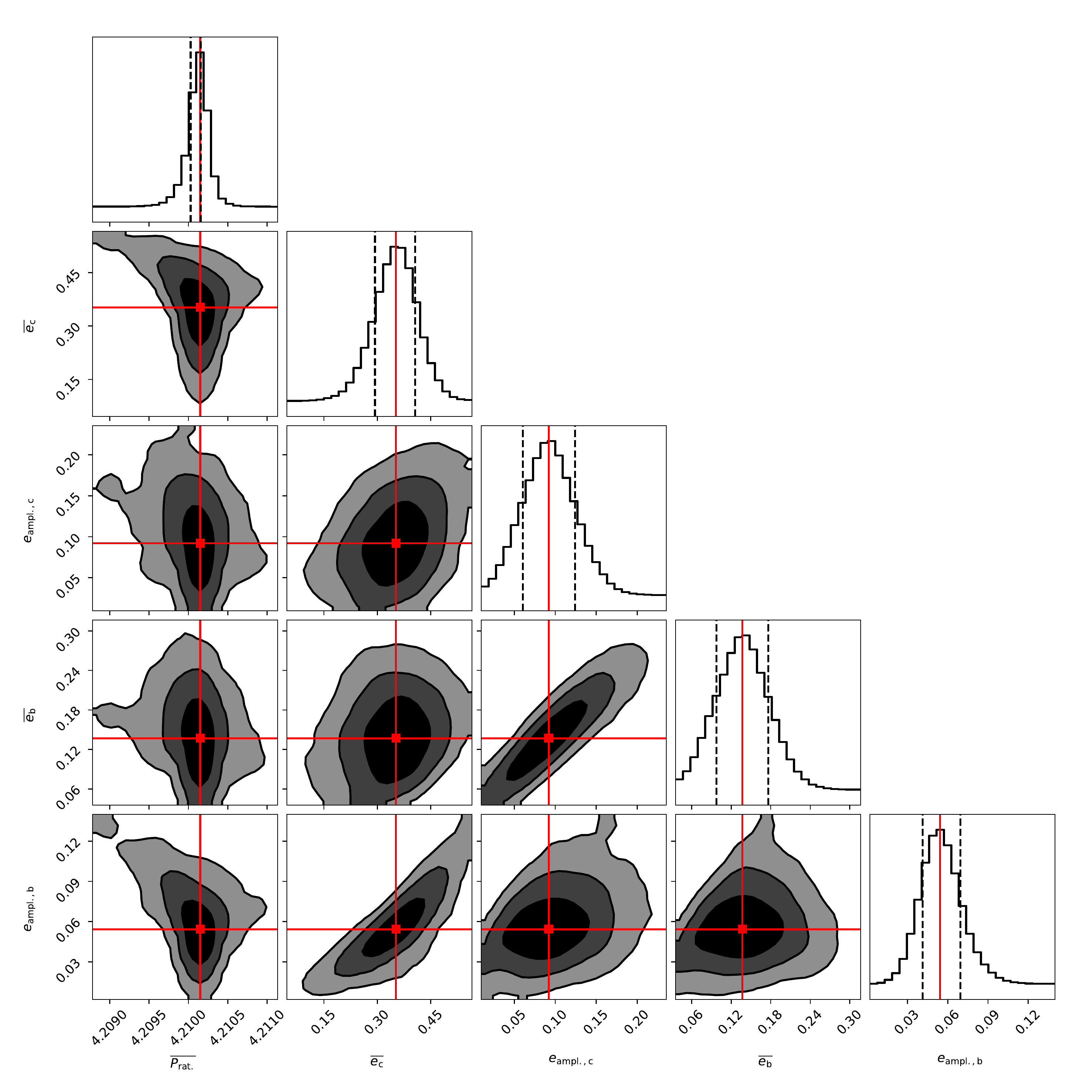}
    \caption{Distribution of the dynamical properties of the MCMC posteriors integrated for 10\,Myr. The two-dimensional contours indicate 1-, 2- and 3-$\sigma$ confidence intervals of the derived posterior distribution. The red crosses indicate the median of the posterior probability distribution. Top to bottom, and left to right: Mean orbital period ratio of the planets $P_{\mathrm{rat.}}$, mean planetary eccentricities and eccentricity amplitudes $\langle e_{\mathrm{c}}\rangle$, $e_{\mathrm{ampl., c}}$, $\langle e_{\mathrm{b}}\rangle$ and $e_{\mathrm{ampl., b}}$, respectively.
    }
    \label{Fig:stabcorner}
\end{figure*}

 \subsection{Dynamical analysis}
 \label{sec:DynamicalAnalysis}

The HD\,107148 b \& c planets orbit their star in close proximity and are consistent with significant eccentricities. Therefore, we find it essential to study the system's long-term stability and dynamical characteristics.
  For our dynamical analysis, we adopted a numerical scheme similar to that used by \citet{Trifonov2020} for the eccentric two-planet system GJ\,1148.
 We performed N-body simulations of the orbital evolution of our adopted best-fit (Model "6") and the achieved MCMC posterior parameter distribution of this model.

We integrated orbits for 10\,Myr using the Wisdom-Holman N-body integrator included in the \texttt{Exo-Striker} \citep[also known as MVS,][]{Wisdom1991}, and a flavor of the  Wisdom-Holman integrator, which can calculate the
General Relativistic precession effects \citep[MVS-GR, see,][]{Trifonov2020}. With the latter integrator, we aimed for a more realistic picture of the long-term evolution of the HD\,107148 system, keeping in mind the moderately close and eccentric orbits of HD\,107148 c \& b, which are affected by GR precession.
We adopted a small numerical time step of 0.2\,d, which we find sufficient for the accurate integration of the system.
While the system is likely very old (see \autoref{table:phys_param}), the 10\,Myr integrations cover $\sim200$ million orbits of the innermost planet HD\,107148 c, which we find adequate for deriving conclusions about the system's long-term stability.

 Similarly to \citet{Trifonov2020}, we monitored the evolution of the planetary
semi-major axes and eccentricities as a function of integration time and rejected orbital configurations as unstable if at any given time $a_b$ or $a_c$ exceeded their starting values by 20\%, or if $e_b$ or $e_c$ became too large and led to crossing orbits.
Since the HD\,107148 b \& c planets are eccentric yet well separated at the period ratio above the 4:1 commensurability, we also monitored the evolution of the period ratio of the planets HD\,107148\,b and c as well as the characteristic resonance angles of the 4:1 MMR configuration;

\begin{eqnarray}
 \theta_1 &=& \lambda_{c} - 4\lambda_{b} + 3\varpi_{b} \\
 \theta_2 &=& \lambda_{c} - 4\lambda_{b} + \varpi_{c} +  2\varpi_{b}\\
 \theta_3 &=& \lambda_{c} - 4\lambda_{b} + 2\varpi_{c} +  \varpi_{b}\\
 \theta_4 &=& \lambda_{c} - 4\lambda_{b} + 3\varpi_{c},
\end{eqnarray}

\noindent
where $\varpi_{\rm b,c}=\Omega_{\rm b,c}+\omega_{\rm b,c}$ are the planetary longitudes of periastron, $\lambda_{\rm b,c}$= M$_{0 \rm b,c}+\varpi_{\rm b,c}$ are the mean longitudes.
Further, we monitored the secular apsidal angle $\Delta\omega$, which is defined as:
\begin{equation}
    \Delta\omega=\omega_{c}-\omega_{b},
\end{equation}
which, if librating around a fixed angle, could indicate that the dynamics of the system is dominated by secular interactions.

 The integration of the best-fit shows that the HD\,107148 system is long-term stable.
 The semi-major axes stay constant over the 10\,Myr of integration time, but we observe a notable osculation of the planetary orbital eccentricities.
 The top panel of \autoref{fig:orbital_evolution} shows the evolution of the eccentricities of the planets using both MVS (solid lines) and MVS-GR (dashed lines).
  The eccentricities can be seen osculating around their mean values of $\overline{e_{b}}=0.14\pm0.04$, $\overline{e_{c}}=0.35\pm0.06$ with different secular times-scales
  for the MVS and the MVS-GR runs. The bottom panel of \autoref{fig:orbital_evolution} shows a GLS periodogram of the eccentricity time series, calculated from a 500\,000 yr evolution, which indicates an osculation with a period of 5707 d for the MVS and a somewhat shorter period near 5581 d for the MVS-GR. For the best fit, we derived a mean period ratio of $P_{\rm rat.}$ = 4.21 and, $\theta_1$, $\theta_2$, $\theta_3$, $\theta_4$, $\Delta\omega$, circulating between 0$^\circ$ and 360$^\circ$. A closer inspection of the trajectory evolution of the resonance angles confirmed that there is no fixed point libration (see \autoref{Fig:resonance}).
  Therefore, we conclude that the best-fit is stable, but not in a 4:1 MMR or involved in secular resonance dynamics. We also conclude that the high-order General Relativistic effects cannot be neglected in the HD\,107148 system for secular time scales. However, they do not affect the system's stability if not included. Therefore, we continued our stability test over the MCMC posterior probability distribution using MVS.\\

 \begin{figure}
    \includegraphics[width=9cm]{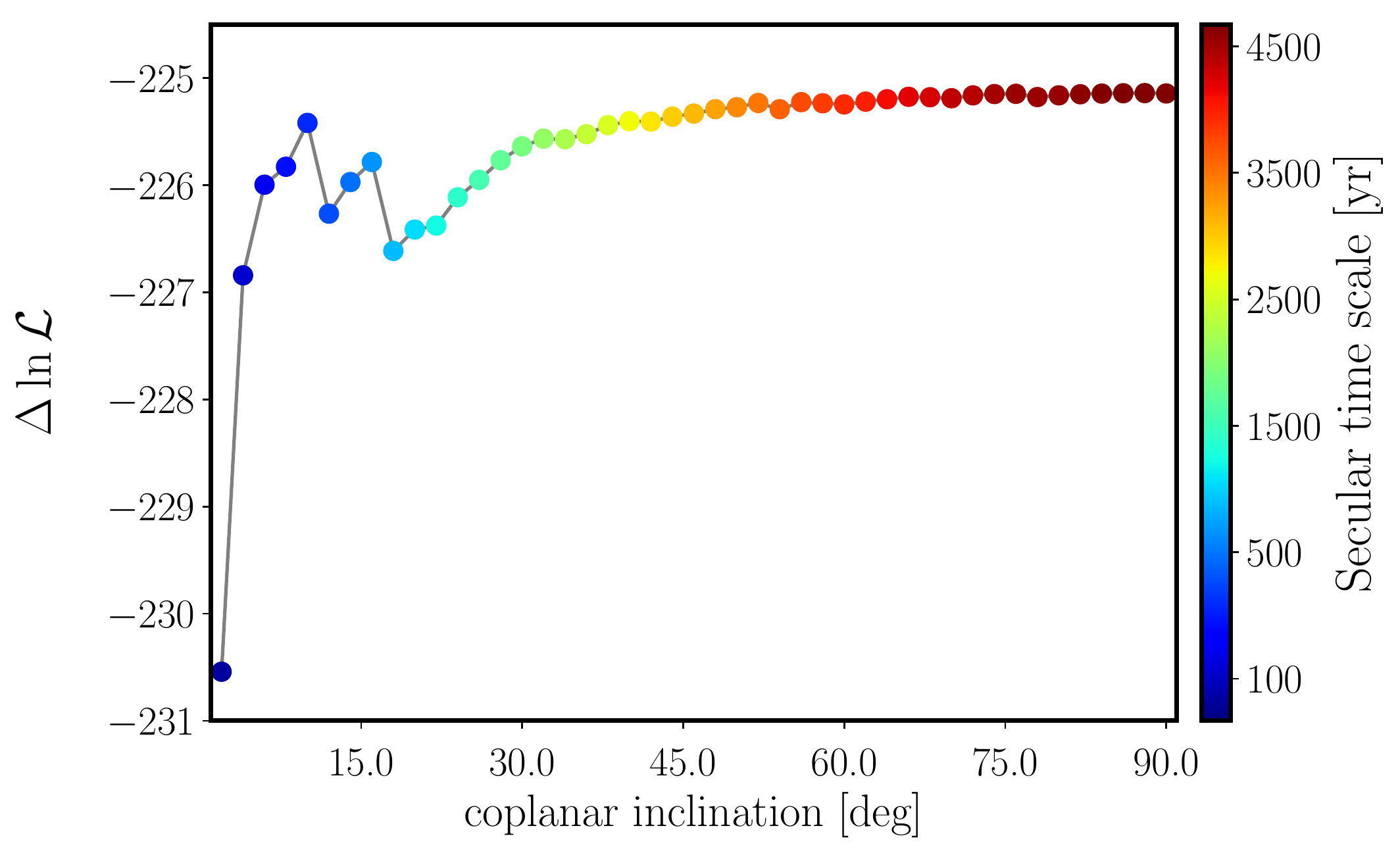}
    \caption{Goodness of coplanar inclined dynamical fits.
All fits are long-term stable, color coded is the secular time scales of the orbital osculation (e.g., see, \autoref{fig:orbital_evolution}).
    }
    \label{Fig:sec_time_incl}
\end{figure}

  To access the overall dynamical properties of the system, we chose to evaluate 10\,000 randomly selected samples from the full MCMC posterior sample (see \autoref{Fig:corner}) and integrated them with the MVS integrator, the same way we did for our best-fit.
  This way, we constructed posteriors of the dynamical parameters of the system.
  In addition, we also constructed posterior distributions of the mean eccentricities, semi-major axes and their amplitudes.

 All examined samples are stable for 10\,Myr, showing similar dynamical behavior as observed in the best-fit N-body test. Thus, we conclude that the HD\,107148 system is generally long-term stable. We find all four resonance angles associated with the 4:1 mean-motion resonant commensurability, $\theta_1$, $\theta_2$, $\theta_3$, $\theta_4$, and the difference of the orbital arguments of periastron (a.k.a. secular apsidal resonance angle) $\Delta\omega=\omega_{\mathrm{c}}-\omega_{\mathrm{b}}$, circulate between 0$^\circ$ and 360$^\circ$, indicating no mean-motion resonant behaviour, as shown in \autoref{Fig:resonance}.

The posterior probability distribution of the dynamically interesting properties of the HD\,107148 system is shown in  \autoref{Fig:stabcorner}. These are the mean
 period ratio of the orbital evolution $P_{\mathrm{rat.}}$, mean planetary eccentricities $\overline{e_{\mathrm{c}}}$, $\overline{e_{\mathrm{b}}}$, and eccentricity amplitudes  $e_{\mathrm{ampl., c}}$, $e_{\mathrm{ampl., b}}$, respectively. The distribution of the mean period ratio shows $P_{\mathrm{rat.}}$ = 4.21$_{-0.03}^{+0.02}$. Generally, the system is consistent with moderate mean eccentricity evolution for both planets; $\overline{e_{\mathrm{c}}}$ = 0.35$_{-0.06}^{+0.05}$, and somewhat lower for the more massive outer planet $\overline{e_{\mathrm{b}}}$ = 0.14$_{-0.04}^{+0.04}$. The eccentricity libration amplitudes drawn from the samples are non-neglegible with $e_{\mathrm{ampl., c}}$ = 0.09$_{-0.03}^{+0.03}$ and $e_{\mathrm{ampl., b}}$ = 0.05$_{-0.01}^{+0.01}$. \autoref{Fig:stabcorner} also shows a clear correlation between  $\overline{e_{\mathrm{b}}}$ and $e_{\mathrm{ampl., c}}$, and $\overline{e_{\mathrm{c}}}$ and $e_{\mathrm{ampl., b}}$.

\subsection{Constraining the dynamical masses.}
 \label{sec:mut_incl}

In an attempt to constrain the dynamical masses of HD\,107148 b and c, we performed N-body modeling adopting different coplanar inclinations.
 We inspect the range of fixed coplanar inclinations from $i$ = 90$^\circ$ to
 2$^\circ$ with a  decreasing step of 2$^\circ$. That is, for each step we fixed; $i_b$ = $i_c$, $\Delta\Omega$ = 0$^{\circ}$, and $\Delta i$ = 0$^{\circ}$, exploring from edge-on to nearly face-on geometries.
 Thus, we keep the system coplanar, but we increase the masses with a factor of $\sim\sin i$,  while optimizing the orbital parameters. This test was made consecutively for better convergence, and the best-fit for $i$ was used as an initial guess for the next fit with lower inclination. For each inclined MLE best-fit, we examine the quality, and the long-term stability in the same ways as we did in \autoref{sec:DynamicalAnalysis}.\looseness=-4

 \autoref{Fig:sec_time_incl} shows the quality of the dynamical fits as a function of the coplanar inclination. The formally best-fit is achieved at edge-on geometry ($i$ = 90$^{\circ}$), for which we achieved similar properties to our two-planet Keplerian (model ``6''). We find that the quality  of the N-body fits in terms of $\Delta\ln\mathcal{L}$ decreases for lower inclination geometries.
Generally, however, all fits up to $i$ = 4$^{\circ}$ are statistically equivalent, whereas the fit with $i$ = 2$^{\circ}$ seems to suggest a significantly worse solution.

In terms of long-term stability, we find that all fits are stable, even for inclinations as low as $i$ = 2$^{\circ}$ leading to planetary masses of m$_b$ = 1.95 $M_{\rm Jup.}$ and m$_c$ = 5.6 $M_{\rm Jup.}$. However, even at these large masses, the HD\,107148 system could still be $\sim6.5$ mutual Hill radii apart, which is above the $\sim3.5$ R$_{\rm Hill,m}$  limit needed for the system to be considered Hill-stable \citep[see,][]{Gladman1993}.
A notable dynamical effect of the larger planetary masses is increasing frequency of the
dynamical secular time scales of the system, which is shown color-coded in  \autoref{Fig:sec_time_incl}.
For $i$ = 90$^{\circ}$ it is about
$\sim5\,600$\,yr (see, e.g., \autoref{fig:orbital_evolution}), but for the nearly face-on configuration with $i$=2$^{\circ}$ it is down to $\sim400$\,yr. The relatively short orbital osculation for large masses within the temporal baseline of the RVs could explain the bad fit quality at low $i$. Therefore, we concluded that our N-body modeling could exclude extremely low inclinations, but overall, we could not constrain the planetary dynamical masses from our RV data. Since orbital inclinations near $i=90^\circ$
are statistically more likely, and our formally best fit is at $i=90^\circ$, we conclude that HD\,107148 is likely on a near coplanar and edge-on configuration.

 \section{Summary and Discussion}
 \label{sec:ConclusionDiscussion}

We present an updated orbital solution for the G-dwarf exoplanet system HD\,107148, and report the detection of a second planet in the system, based on the available archival HIRES data and our newly obtained HARPS data. The original publication of \citet{Butler2006b} reported an exoplanet with a period of $\sim48\,\mathrm{d}$, which has since been revised by \citet{Butler2017} to $\sim 77\,\mathrm{d}$. We find that the literature orbital period of $\sim 48\,\mathrm{d}$ for HD\,107148 is in fact an alias of the $\sim 29\,\mathrm{d}$ period of the lunar cycle and the true $\sim 77\,\mathrm{d}$ signal.
We were able to independently confirm the $77\,\mathrm{d}$ period, as well as to detect an additional $\sim 18\,\mathrm{d}$ signal, which we also interpret to be of planetary nature.
Similar conclusions have been reached independently by \citet{Rosenthal2021}, based
on an extended HIRES RV data set.

  \begin{figure}[ht]
    \includegraphics[width=0.48\textwidth]{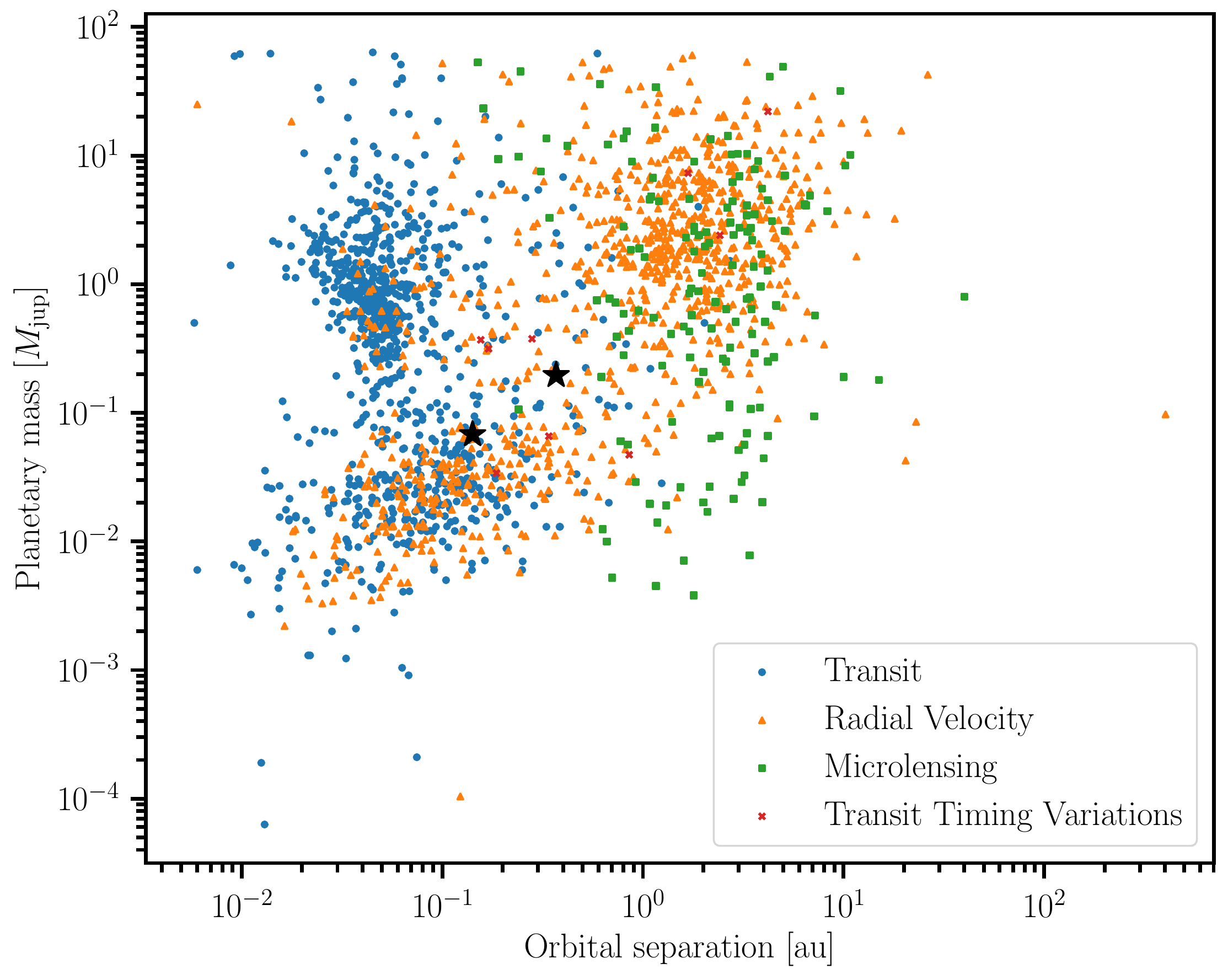}
   \caption{Mass-separation plot for known exoplanets color-coded according to their detection method: {\em blue circles}: Transit, {\em orange triangles}: Radial Velocity, {\em green squares}: Microlensing, {\em red crosses}: Transit Timing Variations. Due to the generally unknown inclination for planets detected via the Radial Velocity method, their masses should be considered as $m\sin{i}$. HD\,107148\,b and c are shown as black stars.
   }
    \label{Fig:demographics}
\end{figure}

  We qualitatively compare different possible orbital configurations for the system, and we reveal the statistically most likely two-planet solution consistent with the available RV data.
  We conclude that the HD\,107148 system consists of a Solar-type star, orbited by two close planets, a Saturn-sized outer planet with an orbital period of $\sim77$\,d and a Neptune-sized inner planet with an orbital period of $\sim18$\,d.
  \autoref{Fig:demographics} shows a mass-separation diagram for known exoplanets\footnote{\url{https://exoplanet.eu}, accessed 25 November 2021}. With minimum planetary masses of $m_b\sin{i}=0.20\,M_{\mathrm{jup}}$, $m_c\sin{i}=0.07\,M_{\mathrm{jup}}$ and orbital separations of $a_b\sin{i}=0.37\,\mathrm{au}$, $a_c\sin{i}=0.14\,\mathrm{au}$, HD\,107148\,b falls into an area sparsely populated with planets (upper black star), as it's less massive than most hot Jupiters and does not orbit as close to its star. HD\,107148\,c falls onto the massive end of the cluster that includes planets detected using transit methods (lower black star). For a Solar-type star like HD\,107148, the two planets are unusually massive and close to their host star. The eccentric configuration of HD\,107148 is likely an aftermath of early planet-planet scattering interactions during, or after, the planetary migration epoch. These physical and orbital characteristics of the HD\,107148 system make it an ideal target for further research about the formation and evolution of planetary systems.

Further, we provide a dynamical orbital evolution and stability analysis of the HD\,107148 system.
 Our extensive N-body simulations yield that the two-planet system configuration is long-term stable within the entire posterior parameter space, and for a large range of co-planar inclinations. We find that the overall dynamics of the system is not consistent with a resonant behavior, but it exhibits large planetary eccentricity osculations with a secular time scale of about 5\,500\,yr for edge-on configuration, and albeit with much lower confidence, it could be down to 400\,yr if the systems' orbit is nearly face-on to the line of sight with Earth.  While our fitting results favor a configuration with both planets on eccentric orbits, the evidence is not strong enough to rule out the case of HD\,107148\,b's orbit being nearly circular at present times. Nevertheless, we find this unlikely since the system is dynamically very active, and the exchange of energy and momentum would periodically trigger eccentricity exchange on secular time scales.

  What makes this system particularly interesting is the fact that it also includes  a white dwarf companion, orbiting the main-sequence star HD\,107148\,A with a projected orbital separation of more than 1\,000 au.
  While the WD companion might only have a small gravitational influence on the planetary system, the system is likely to have been covered in a planetary  nebula.  HD\,107148 is one of only a handful known exoplanet systems featuring a WD stellar component. Some of these include the systems; HD\,13445 \citep[Gl\,86,][]{Queloz2000, Els2001, Lagrange2006}, HD\,27442 \citep{Butler2001, Chauvin2006, Mugrauer2007}, HIP\,116454 \citep{Vanderburg2015}, TOI\,1259 \citep{Martin2021} and HD\,147513 \citep{Mayor2004, AlexanderLourens1969} \citep[for a more detailed list, see][]{Martin2021}. Assuming that both the WD and the main sequence star have formed at the same time, the age of the system can be constrained from the total age of the WD, which is the sum of its cooling age and the age of its progenitor. However, the age of the HD\,107148 system estimated this way is poorly constrained, future observations could help better constraining the mass and cooling age of the white dwarf, allowing for better estimates of the systems age.

Based on observations collected at the European Organization for Astronomical Research in
the Southern Hemisphere under ESO programmes: 072.C-0488, 183.C-0972, 097.C-0090, 0100.C-0414, 60.A-9700, 0101.C-0232. Based on observations collected at the Calar Alto (CAHA, Almer\'{\i}a, Spain) under program: F17-3.5-019.
This research has made use of the SIMBAD database, operated at CDS, Strasbourg, France.
This work has made use of data from the European Space Agency (ESA)
mission {\it Gaia} (\url{https://www.cosmos.esa.int/gaia}), processed by
the {\it Gaia} Data Processing and Analysis Consortium (DPAC,
\url{https://www.cosmos.esa.int/web/gaia/dpac/consortium}). Funding
for the DPAC has been provided by national institutions, in particular
the institutions participating in the {\it Gaia} Multilateral Agreement.
T.T. and M. K. acknowledge support by the DFG Research Unit FOR 2544 ”Blue Planets around Red Stars” project No. KU 3625/2-1.
Th. H.\ acknowledges support by the DFG Research Unit FOR~2544 {\it Blue  Planets around Red Stars}. Th. H.\ further acknowledges support by the DFG Priority Program SPP~1992
{\it Exploring the Diversity of Extrasolar Planets}.
M.H.L. is supported in part by Hong Kong RGC grant HKU 17305618.

\facilities{ESO-3.6m/HARPS, KECK/HIRES, Calar Alto-3.5m/CARMENES}

\software{
          Exo-Striker~\citep{Trifonov2019_es},
          SERVAL~\citep{Zechmeister2018},
          }

\clearpage

\clearpage

\bibliography{bibliography}{}
\bibliographystyle{aasjournal}

\begin{appendix} 

\label{appendix}

In this Appendix, we show additional plots, including the HIRES window function, trajectory plots of the resonance angles and the corner plot for our MCMC analysis, as well as tables containing our HARPS and CARMENES data.
 \autoref{Fig:HIRESwindow} shows the window function of the HIRES data of HD\,107148 provided by \cite{TalOr2018}.
 \autoref{Fig:resonance} shows the trajectory plots of the evolution of the four resonance angles associated with a 4:1 mean motion resonance.
 \autoref{Fig:corner} shows the posterior distribution of our MCMC sample of the combined HARPS and HIRES data of HD\,107148.
 \autoref{tab:HARPS_1} and \autoref{tab:HARPS_2} show the HARPS radial velocity and activity index measurements for HD\,107148, derived with the DRS and Serval pipelines.
 \autoref{tab:CARM_VIS} and \autoref{tab:CARM_NIR} show the CARMENES radial velocity and activity index measurements in the visual and near-infrared range, derived with the Serval pipeline.

 \setcounter{table}{0}
\renewcommand{\thetable}{A\arabic{table}}

\setcounter{figure}{0}
\renewcommand{\thefigure}{A\arabic{figure}}

\begin{figure*}[ht]
    \includegraphics[width=\textwidth]{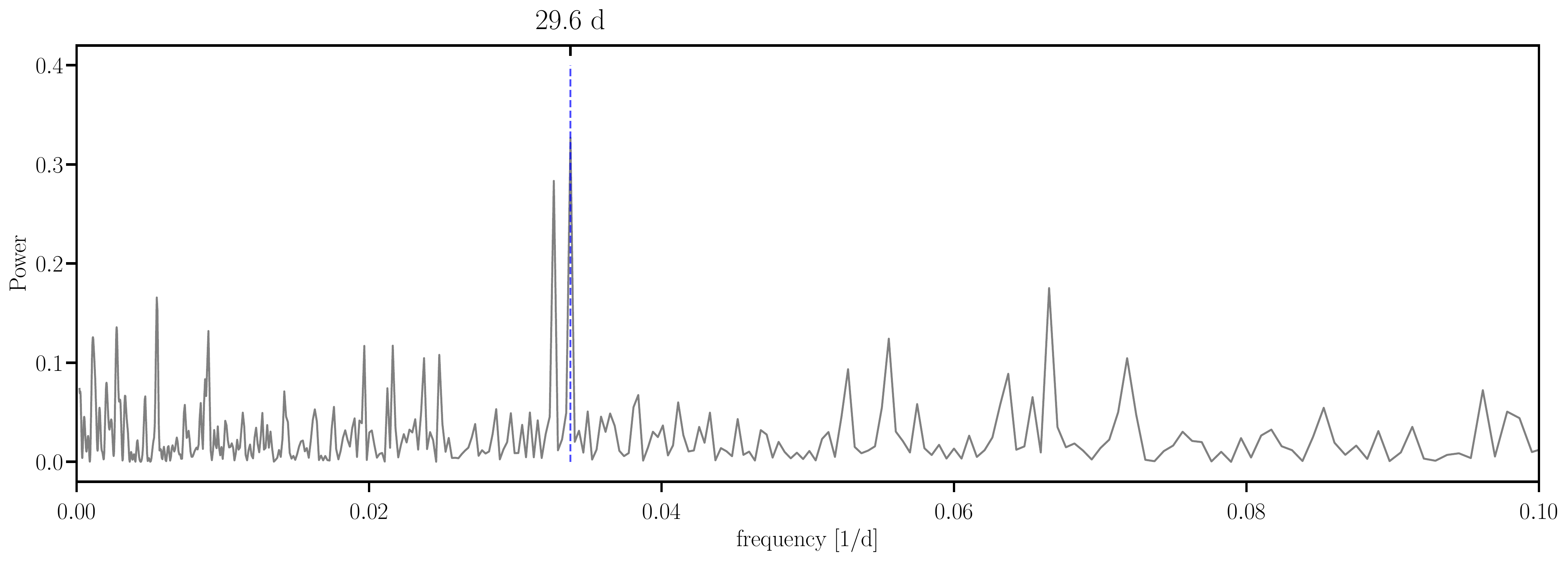}
   \caption{Window function of the HIRES data set. The highest peak at a period of $\sim 29.6\,\mathrm{d}$ corresponds to the lunar cycle, which affects the Keck-HIRES observational schedule.}
    \label{Fig:HIRESwindow}
\end{figure*}

\begin{figure*}[ht]
    \includegraphics[width=0.23\textwidth]{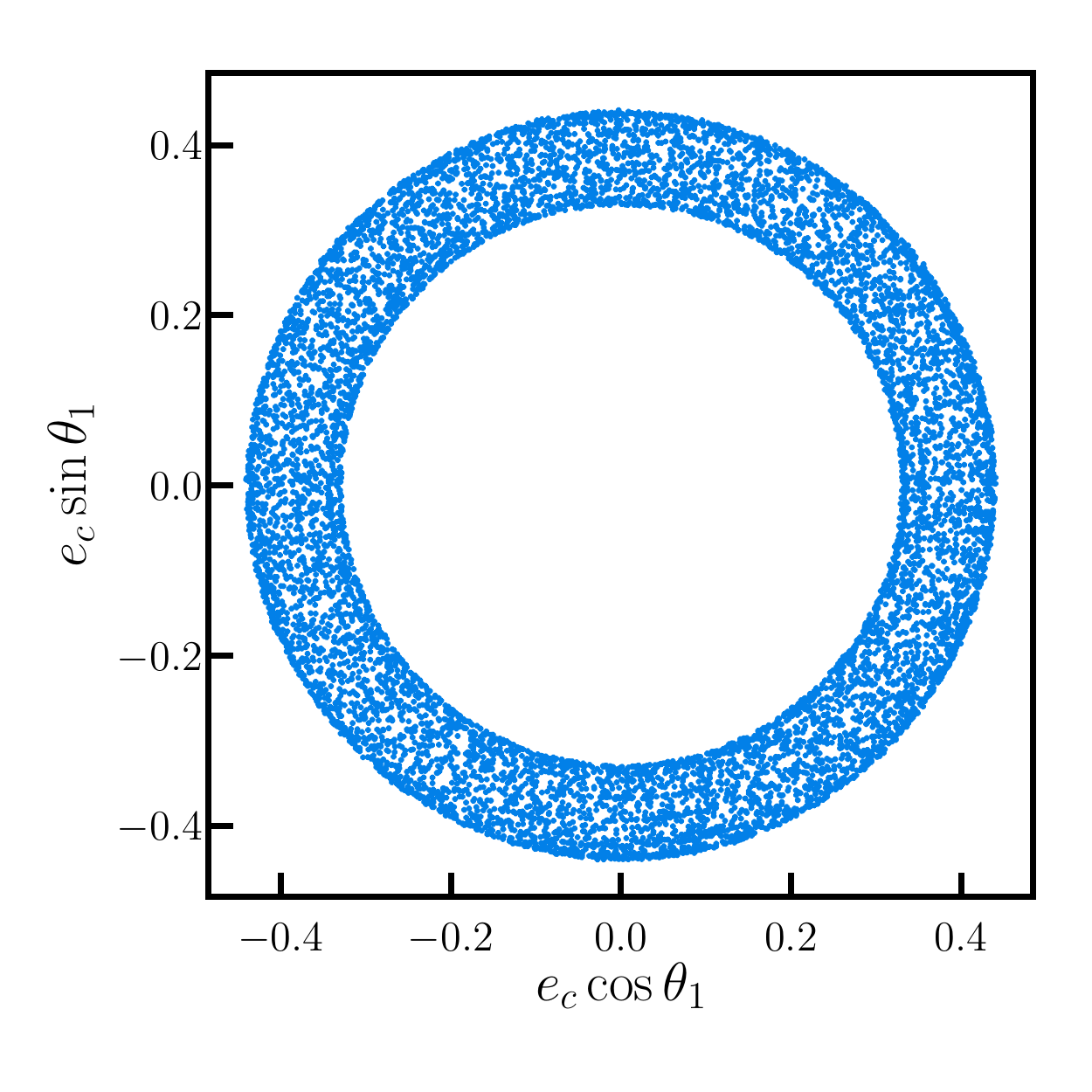}
    \includegraphics[width=0.23\textwidth]{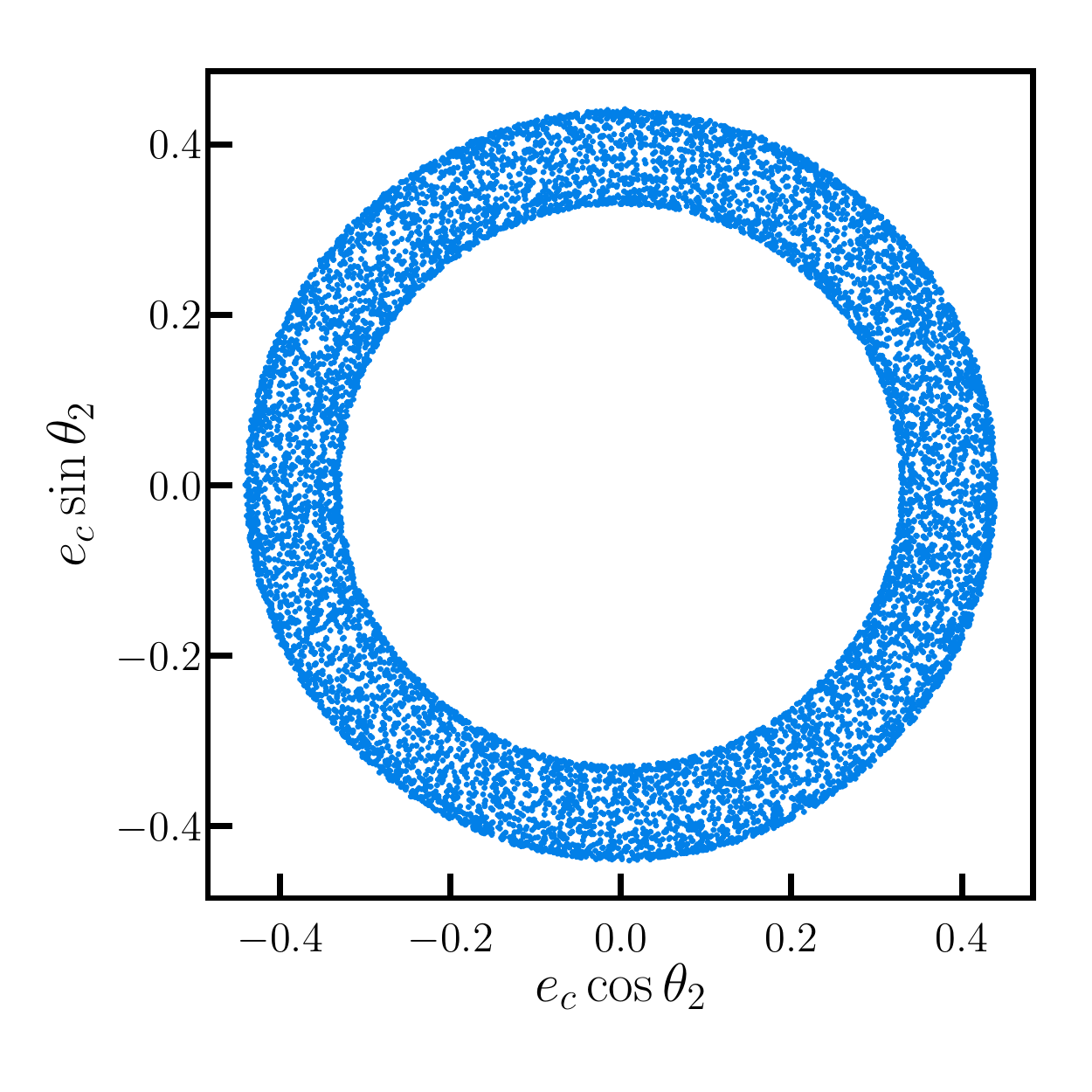}
    \includegraphics[width=0.23\textwidth]{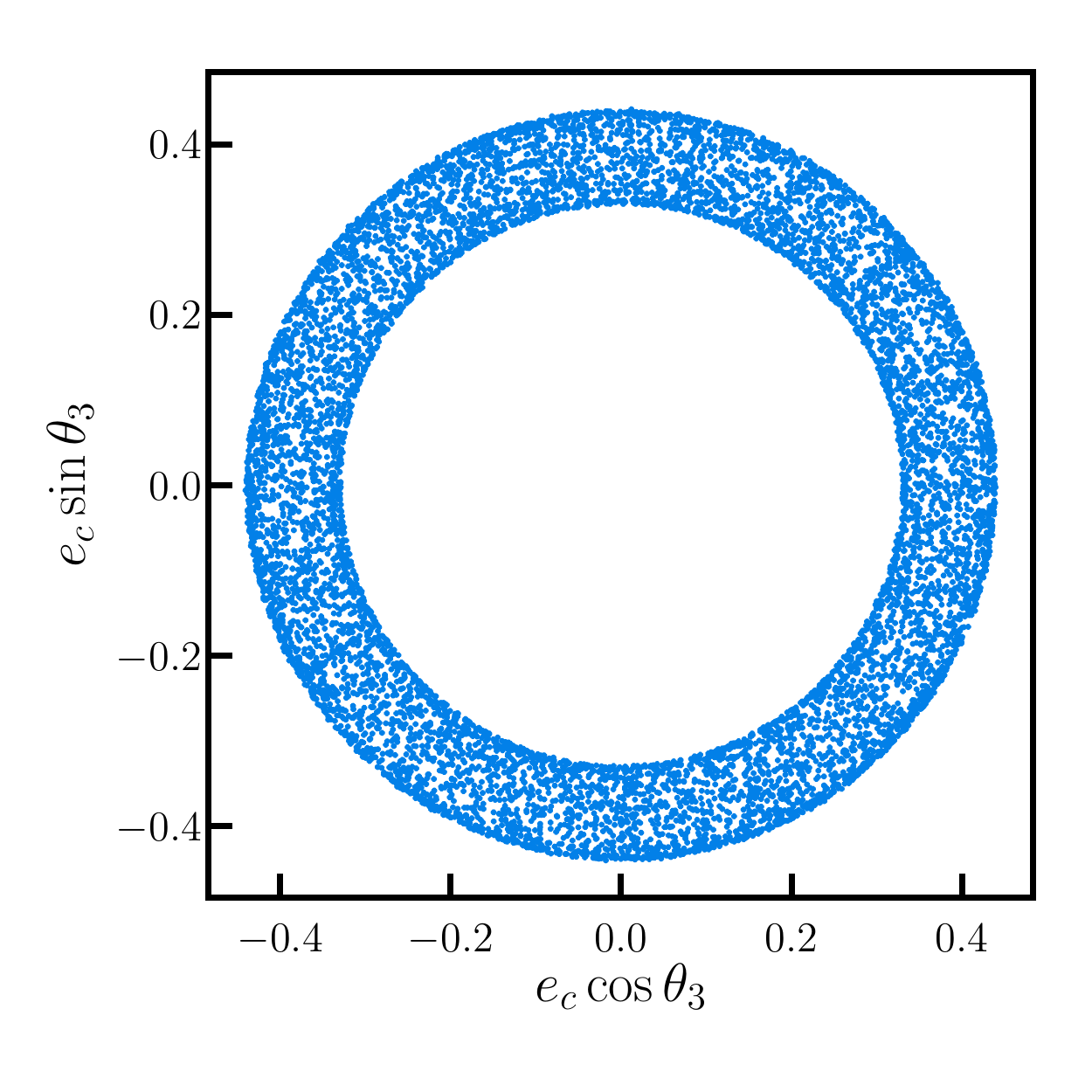}
    \includegraphics[width=0.23\textwidth]{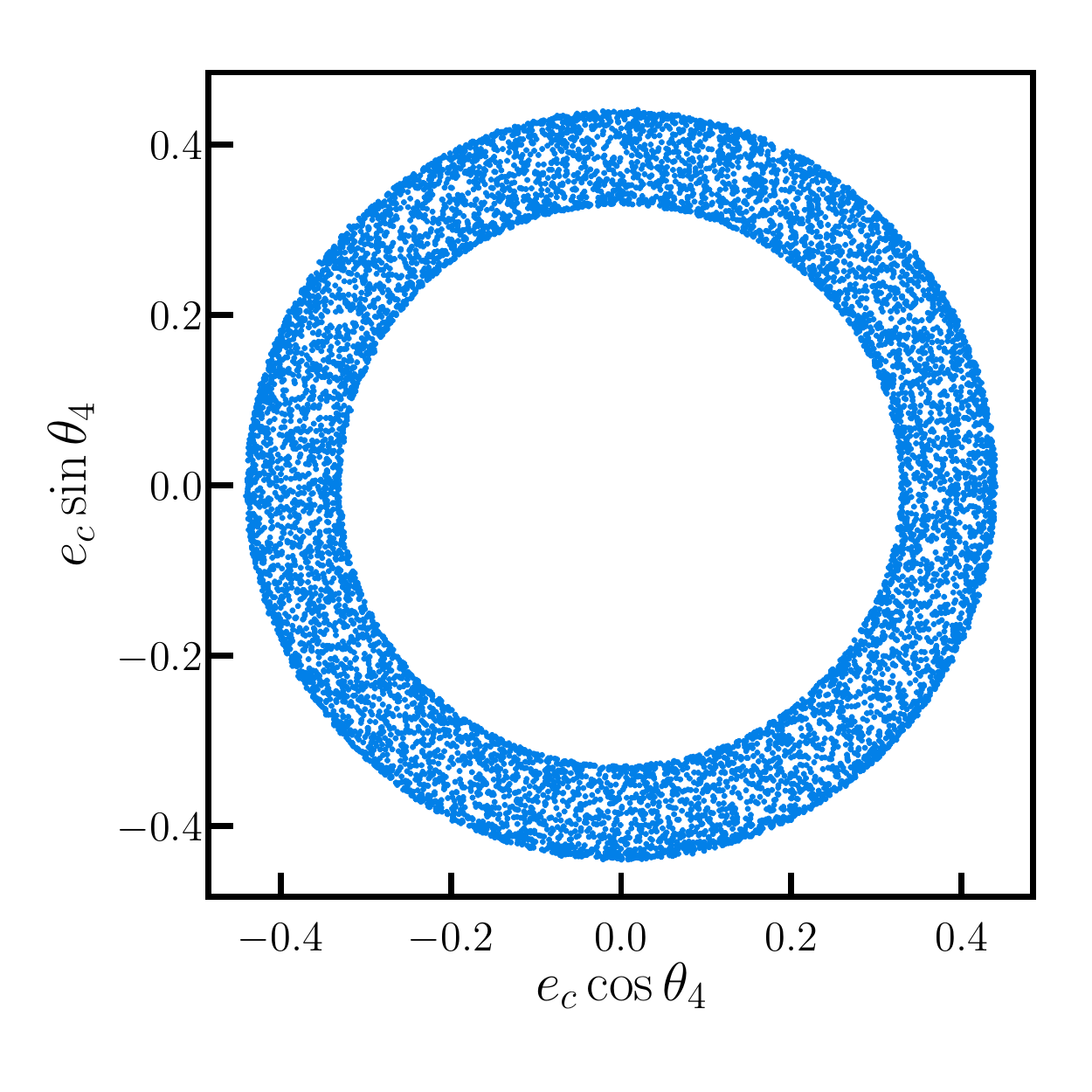}\\
    \includegraphics[width=0.23\textwidth]{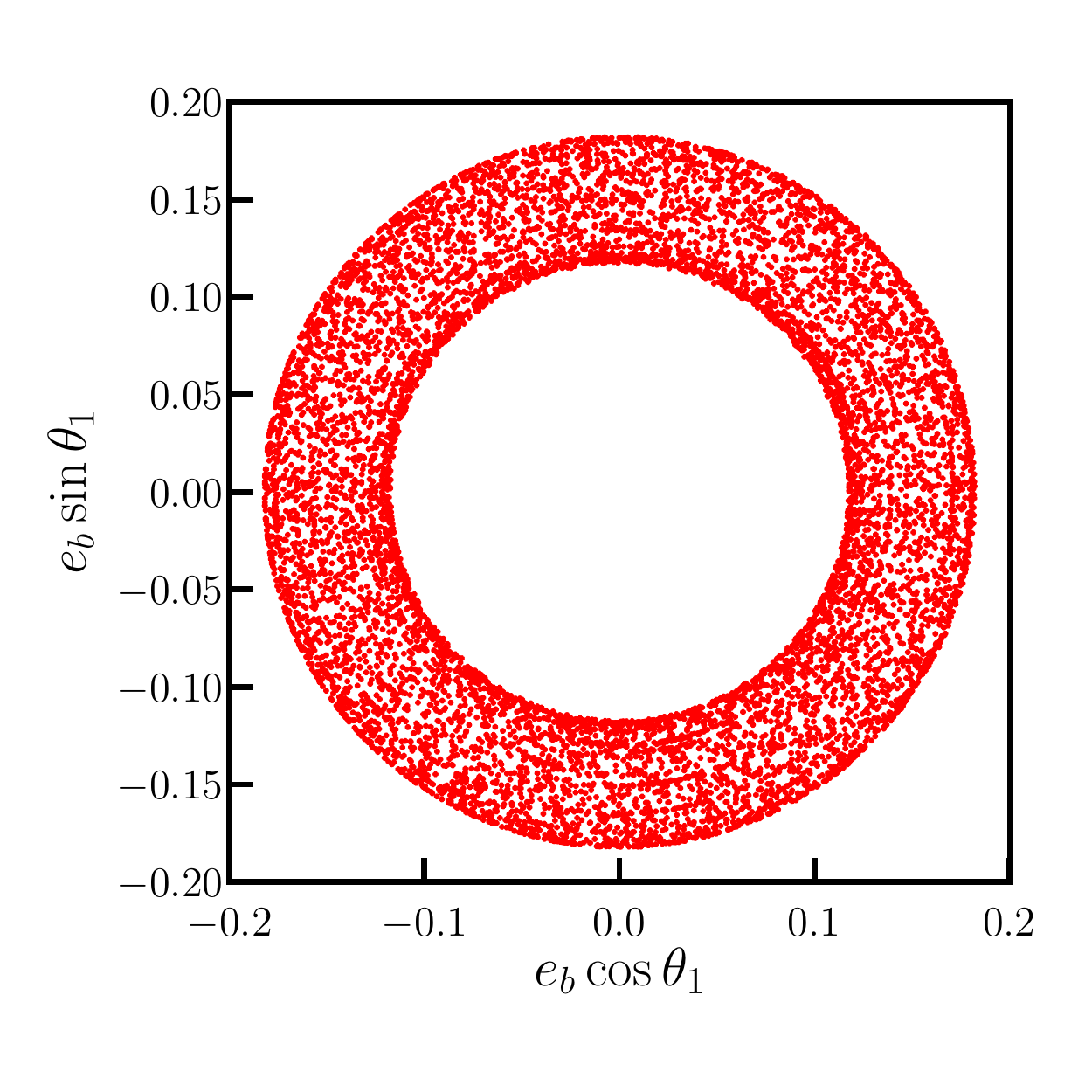}
    \includegraphics[width=0.23\textwidth]{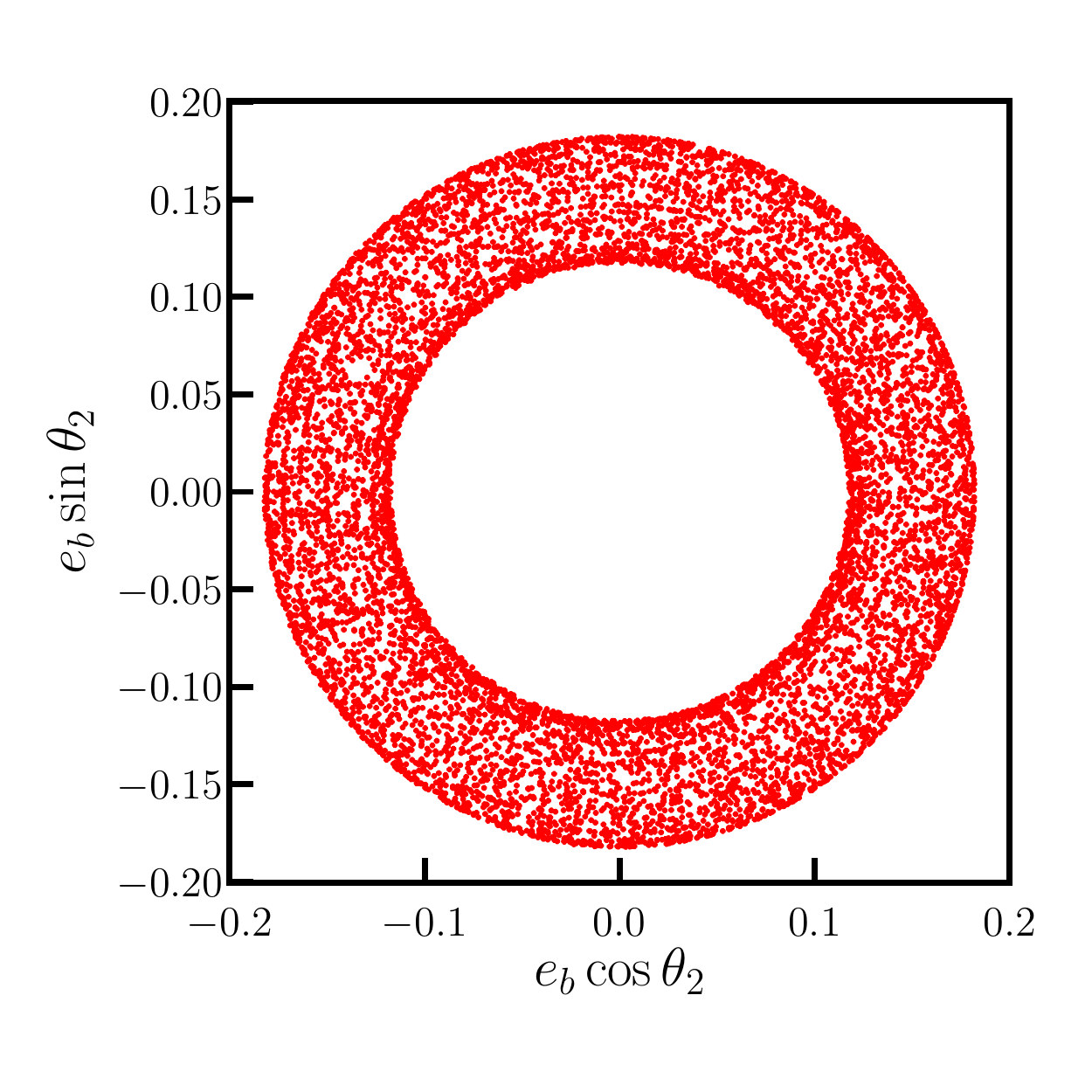}
    \includegraphics[width=0.23\textwidth]{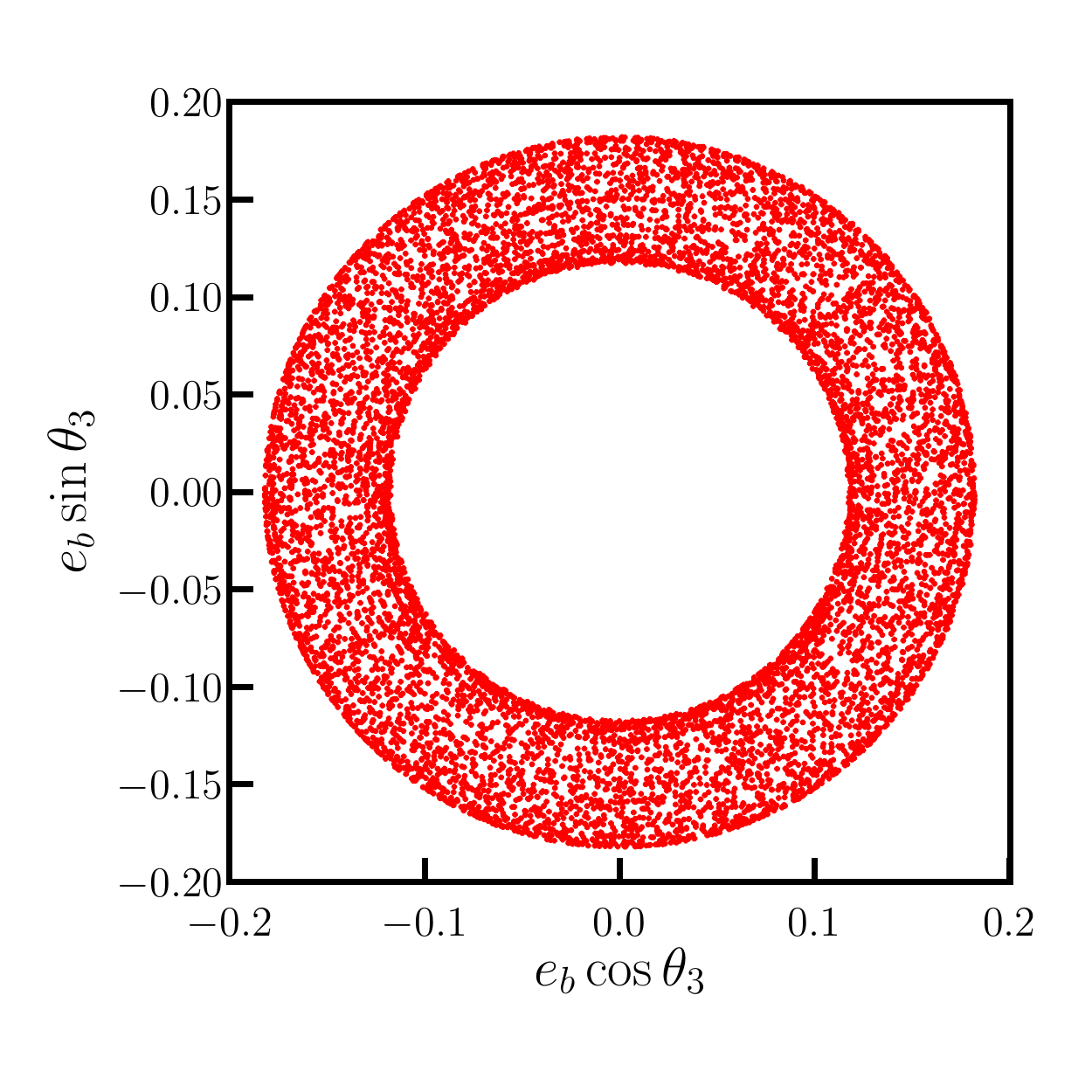}
    \includegraphics[width=0.23\textwidth]{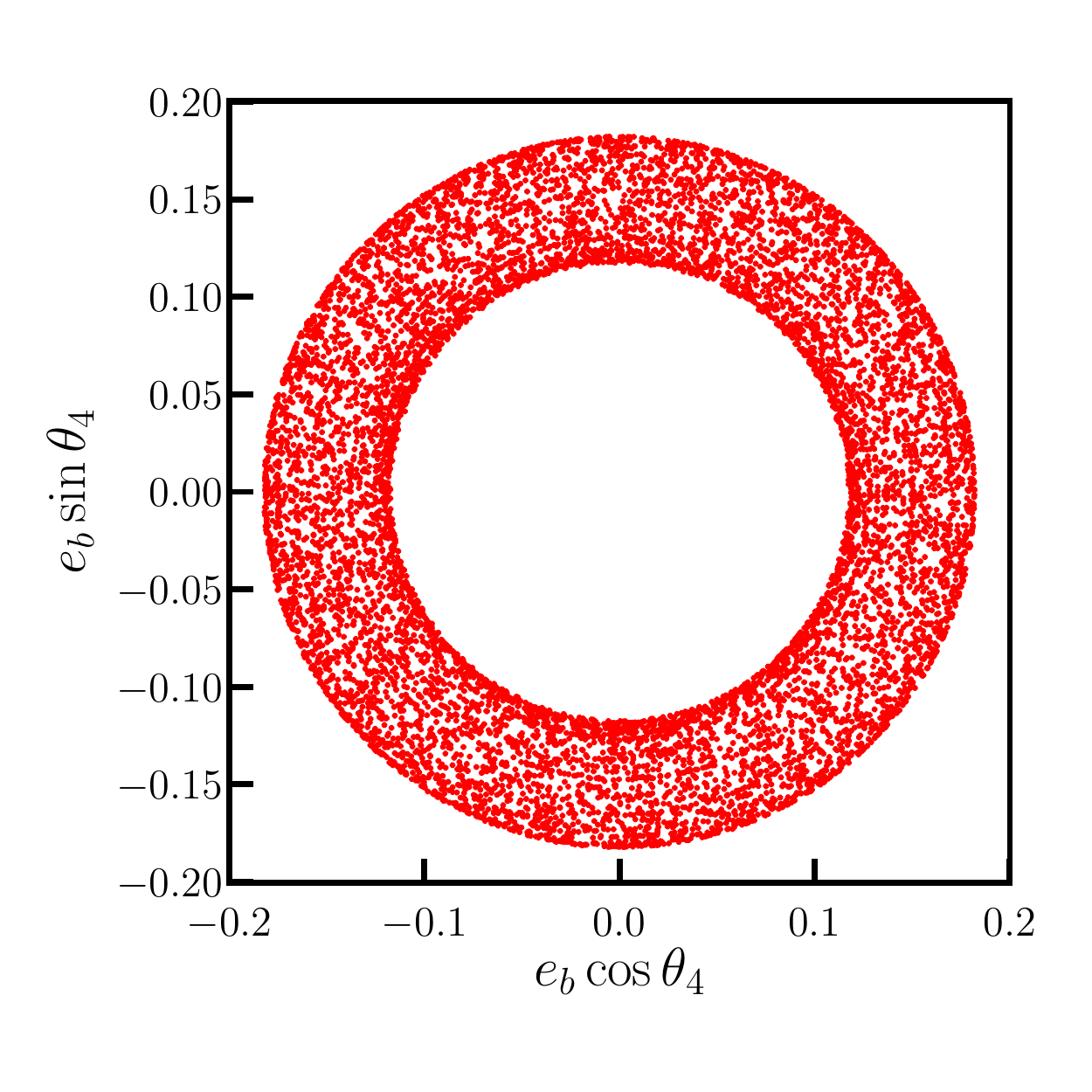}\\
    \caption{Trajectory plots for the resonance angles $\theta_1$, $\theta_2$, $\theta_3$ and $\theta_4$ for HD\,107148\,b ({\em red}) and HD\,107148\,c ({\em blue}).}
    \label{Fig:resonance}
\end{figure*}

    \begin{figure*}[ht]
    \includegraphics[width=18cm]{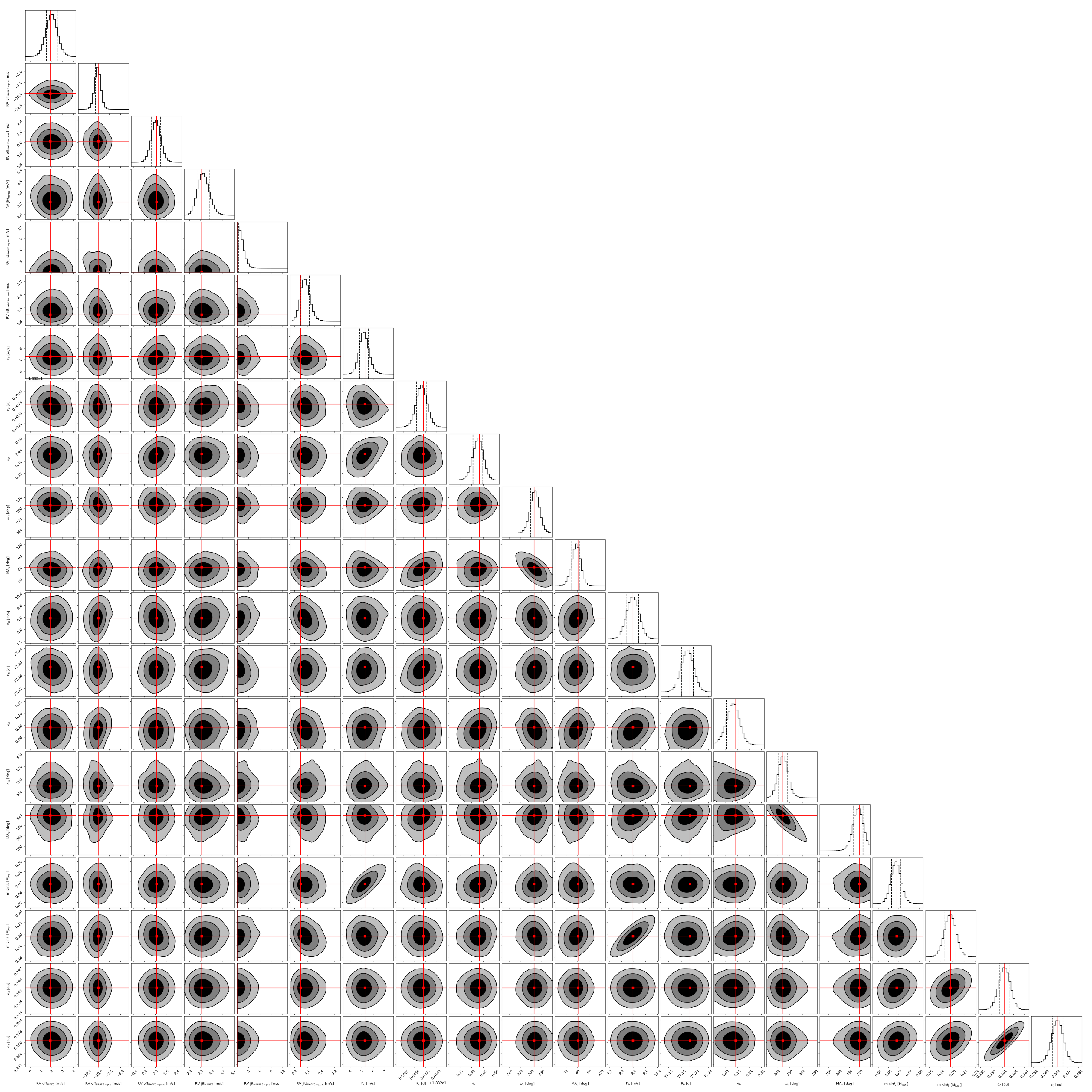}
    \caption{Correlations of the parameter posteriors achieved from the MCMC analysis. Red intersections denote the position of the best-fit values. The two-dimensional contours indicate 1-, 2- and 3-$\sigma$ confidence intervals of the posterior distribution. {\em Top to bottom, and left to right}: RV offsets of HIRES and HARPS (pre- and post- fiber upgrade), RV jitters of HIRES and HARPS, $K_{\mathrm{c}}$, $P_{\mathrm{c}}$, $e_{\mathrm{c}}$, $\omega_{\mathrm{c}}$, $M_{\mathrm{c}}$, $K_{\mathrm{b}}$, $P_{\mathrm{b}}$, $e_{\mathrm{b}}$, $\omega_{\mathrm{b}}$, $M_{\mathrm{b}}$, $m_{\mathrm{c}}\sin{i}$, $m_{\mathrm{b}}\sin{i}$, $a_{\mathrm{c}}$, $a_{\mathrm{b}}$.}
    \label{Fig:corner}
\end{figure*}

\begin{table*}
\caption{HARPS Doppler measurements and activity index measurements of HD\,107148, derived with SERVAL} 
\label{tab:HARPS_1}

\centering  
\resizebox{\textwidth}{!}{
\begin{tabular}{c c c c c c c c c c c c c c c c c c c} 

\hline\hline    
\noalign{\vskip 0.5mm}

Epoch & RV & $\sigma_{RV}$  &  CRX & $\sigma_{\rm CRX}$ &  dLW & $\sigma_{\rm dLW} $  &  H$\alpha$ & $\sigma_{H\alpha}$  &  NaD$_1$ & $\sigma_{\rm NaD_1}$  &  NaD$_2$ & $\sigma_{\rm NaD_2}$  &  FWHM & $\sigma_{\rm FWHM} $  &  Cont & $\sigma_{\rm Cont}$  &  BIS & $\sigma_{\rm BIS} $  \\ 

[JD] & [m\,s$^{-1}$] &  [m\,s$^{-1}$]  &  [m\,s$^{-1}$] &  [m\,s$^{-1}$]  &  [m\,s$^{-1}$ * km\,s$^{-1}$]& [m\,s$^{-1}$ * km\,s$^{-1}$]  &   &   &  &   &   &   &  [km\,s$^{-1}$]& [km\,s$^{-1}$]  &   &   &   [km\,s$^{-1}$]&  [km\,s$^{-1}$]  \\ 

\hline     
\noalign{\vskip 0.5mm}

2453410.74594   &   -20.18   &   1.13 &   7.55   &   11.25 &   -52.50   &   4.15 &   0.4294   &   0.0012 &   0.2858   &   0.0015   &   0.3889   &   0.0020   &   7.151   &   0.072   &   53.21   &   0.53   &   -28.54   &   -0.29   \\  2453412.76455   &   -20.14   &   1.26 &   0.03   &   11.76 &   -56.90   &   4.25 &   0.4284   &   0.0013 &   0.2837   &   0.0017   &   0.3858   &   0.0022   &   7.156   &   0.072   &   53.25   &   0.53   &   -27.49   &   -0.27   \\  2453785.77908   &   -13.27   &   1.76 &   21.94   &   16.40 &   -35.70   &   4.94 &   0.4318   &   0.0019 &   0.2872   &   0.0025   &   0.4025   &   0.0032   &   7.165   &   0.072   &   52.94   &   0.53   &   -31.45   &   -0.31   \\  2453789.82758   &   -19.10   &   1.63 &   3.50   &   14.15 &   -36.71   &   4.62 &   0.4286   &   0.0017 &   0.2876   &   0.0022   &   0.3949   &   0.0028   &   7.162   &   0.072   &   52.99   &   0.53   &   -27.70   &   -0.28   \\  2455291.68074   &   0.11   &   1.05 &   -22.06   &   11.40 &   -30.42   &   4.30 &   0.4244   &   0.0011 &   0.2908   &   0.0014   &   0.3823   &   0.0018   &   7.158   &   0.072   &   52.90   &   0.53   &   -24.67   &   -0.25   \\  2455291.68331   &   1.88   &   0.99 &   11.88   &   10.46 &   -29.80   &   4.55 &   0.4249   &   0.0010 &   0.2908   &   0.0013   &   0.3835   &   0.0017   &   7.165   &   0.072   &   52.88   &   0.53   &   -22.14   &   -0.22   \\  2455291.68564   &   -0.62   &   1.08 &   -8.87   &   13.25 &   -26.48   &   4.25 &   0.4210   &   0.0011 &   0.2926   &   0.0014   &   0.3816   &   0.0019   &   7.167   &   0.072   &   52.88   &   0.53   &   -30.23   &   -0.30   \\  2455291.68824   &   3.43   &   1.04 &   -2.61   &   11.36 &   -27.51   &   4.73 &   0.4247   &   0.0011 &   0.2912   &   0.0014   &   0.3828   &   0.0018   &   7.163   &   0.072   &   52.86   &   0.53   &   -24.36   &   -0.24   \\  2455300.62303   &   -7.00   &   0.55 &   -8.60   &   7.00 &   -28.15   &   4.07 &   0.4244   &   0.0006 &   0.2860   &   0.0007   &   0.3873   &   0.0009   &   7.164   &   0.072   &   52.92   &   0.53   &   -28.49   &   -0.28   \\  2455300.71809   &   -6.84   &   0.45 &   -15.13   &   6.83 &   -28.62   &   3.80 &   0.4235   &   0.0005 &   0.2915   &   0.0006   &   0.3763   &   0.0008   &   7.167   &   0.072   &   52.92   &   0.53   &   -29.11   &   -0.29   \\  2455308.63859   &   0.60   &   0.38 &   -10.61   &   8.04 &   -22.02   &   3.34 &   0.4230   &   0.0004 &   0.2870   &   0.0005   &   0.3848   &   0.0006   &   7.164   &   0.072   &   52.89   &   0.53   &   -23.55   &   -0.24   \\  2455322.64701   &   -16.58   &   0.45 &   -23.44   &   7.77 &   -23.96   &   3.62 &   0.4245   &   0.0004 &   0.2923   &   0.0006   &   0.3885   &   0.0007   &   7.170   &   0.072   &   52.88   &   0.53   &   -27.36   &   -0.27   \\  2455342.62344   &   -17.67   &   0.50 &   -15.81   &   7.31 &   -25.33   &   3.45 &   0.4260   &   0.0005 &   0.2914   &   0.0006   &   0.3889   &   0.0008   &   7.170   &   0.072   &   52.89   &   0.53   &   -25.94   &   -0.26   \\  2457559.61595   &   -1.92   &   0.68 &   -8.10   &   9.13 &   2.96   &   1.62 &   0.4299   &   0.0006 &   0.2905   &   0.0008   &   0.3850   &   0.0010   &   7.193   &   0.072   &   52.67   &   0.53   &   -18.78   &   -0.19   \\  2457559.62209   &   -1.32   &   0.69 &   4.52   &   8.45 &   6.29   &   1.56 &   0.4315   &   0.0006 &   0.2930   &   0.0008   &   0.3851   &   0.0011   &   7.194   &   0.072   &   52.66   &   0.53   &   -17.84   &   -0.18   \\  2457559.62829   &   -1.52   &   0.69 &   1.01   &   8.58 &   7.50   &   1.45 &   0.4297   &   0.0006 &   0.2927   &   0.0008   &   0.3846   &   0.0010   &   7.194   &   0.072   &   52.66   &   0.53   &   -16.54   &   -0.17   \\  2457560.51145   &   0.64   &   0.47 &   11.57   &   7.05 &   3.22   &   0.99 &   0.4315   &   0.0005 &   0.2905   &   0.0006   &   0.3787   &   0.0008   &   7.195   &   0.072   &   52.70   &   0.53   &   -15.25   &   -0.15   \\  2457560.51748   &   1.13   &   0.48 &   12.71   &   7.05 &   5.63   &   1.07 &   0.4307   &   0.0005 &   0.2920   &   0.0006   &   0.3806   &   0.0008   &   7.194   &   0.072   &   52.69   &   0.53   &   -16.61   &   -0.17   \\  2457560.52374   &   1.52   &   0.51 &   8.05   &   7.45 &   2.56   &   1.07 &   0.4309   &   0.0005 &   0.2912   &   0.0006   &   0.3787   &   0.0008   &   7.194   &   0.072   &   52.71   &   0.53   &   -15.34   &   -0.15   \\  2457609.45890   &   6.20   &   0.60 &   8.18   &   9.26 &   8.34   &   1.46 &   0.4295   &   0.0006 &   0.2916   &   0.0007   &   0.3812   &   0.0009   &   7.192   &   0.072   &   52.66   &   0.53   &   -15.09   &   -0.15   \\  2457609.46470   &   8.48   &   0.60 &   6.63   &   8.66 &   3.18   &   1.37 &   0.4293   &   0.0006 &   0.2913   &   0.0007   &   0.3807   &   0.0009   &   7.188   &   0.072   &   52.72   &   0.53   &   -16.52   &   -0.17   \\  2457609.47102   &   7.16   &   0.63 &   8.37   &   8.89 &   4.27   &   1.43 &   0.4289   &   0.0006 &   0.2919   &   0.0008   &   0.3793   &   0.0010   &   7.193   &   0.072   &   52.69   &   0.53   &   -15.01   &   -0.15   \\  2458082.86206   &   4.21   &   0.78 &   -9.27   &   7.85 &   7.37   &   1.78 &   0.4269   &   0.0008 &   0.2924   &   0.0010   &   0.3954   &   0.0013   &   7.193   &   0.072   &   52.69   &   0.53   &   -17.15   &   -0.17   \\  2458082.86809   &   3.34   &   0.71 &   -2.40   &   7.66 &   14.87   &   1.80 &   0.4276   &   0.0007 &   0.3145   &   0.0009   &   0.4083   &   0.0012   &   7.191   &   0.072   &   52.64   &   0.53   &   -13.42   &   -0.13   \\  2458082.87412   &   8.49   &   0.79 &   -21.46   &   7.16 &   50.38   &   3.14 &   0.4275   &   0.0008 &   0.3572   &   0.0011   &   0.4405   &   0.0014   &   7.189   &   0.072   &   52.29   &   0.52   &   -17.25   &   -0.17   \\  2458144.78547   &   4.92   &   0.52 &   2.09   &   6.35 &   10.46   &   1.22 &   0.4259   &   0.0005 &   0.2813   &   0.0006   &   0.3865   &   0.0008   &   7.189   &   0.072   &   52.69   &   0.53   &   -15.22   &   -0.15   \\  2458144.79145   &   3.98   &   0.53 &   -0.15   &   6.23 &   6.59   &   1.38 &   0.4263   &   0.0006 &   0.2804   &   0.0007   &   0.3852   &   0.0009   &   7.188   &   0.072   &   52.70   &   0.53   &   -14.17   &   -0.14   \\  2458144.79767   &   3.93   &   0.54 &   6.28   &   7.18 &   7.82   &   1.24 &   0.4272   &   0.0006 &   0.2798   &   0.0007   &   0.3852   &   0.0009   &   7.190   &   0.072   &   52.71   &   0.53   &   -16.48   &   -0.16   \\  2458145.80913   &   7.11   &   0.65 &   -9.98   &   7.14 &   8.02   &   1.42 &   0.4256   &   0.0007 &   0.2766   &   0.0008   &   0.3973   &   0.0010   &   7.189   &   0.072   &   52.70   &   0.53   &   -11.90   &   -0.12   \\  2458145.81534   &   6.58   &   0.63 &   -12.52   &   6.56 &   10.26   &   1.40 &   0.4265   &   0.0006 &   0.2741   &   0.0008   &   0.3954   &   0.0010   &   7.189   &   0.072   &   52.69   &   0.53   &   -16.07   &   -0.16   \\  2458145.82144   &   5.38   &   0.65 &   -5.41   &   7.48 &   9.38   &   1.42 &   0.4264   &   0.0006 &   0.2761   &   0.0008   &   0.3934   &   0.0010   &   7.186   &   0.072   &   52.70   &   0.53   &   -14.57   &   -0.15   \\  2458171.73165   &   -0.12   &   0.78 &   9.64   &   11.08 &   9.11   &   1.85 &   0.4285   &   0.0009 &   0.2843   &   0.0009   &   0.3954   &   0.0011   &   7.291   &   0.073   &   51.84   &   0.52   &   -31.21   &   -0.31   \\  2458171.73797   &   0.05   &   0.82 &   -3.06   &   10.94 &   6.30   &   2.08 &   0.4264   &   0.0009 &   0.2830   &   0.0009   &   0.3949   &   0.0012   &   7.296   &   0.073   &   51.89   &   0.52   &   -32.95   &   -0.33   \\  2458171.74417   &   0.81   &   0.83 &   5.04   &   11.41 &   7.23   &   1.80 &   0.4266   &   0.0009 &   0.2820   &   0.0010   &   0.3937   &   0.0012   &   7.285   &   0.073   &   51.95   &   0.52   &   -26.46   &   -0.26   \\  2458195.77177   &   -13.79   &   0.55 &   5.68   &   7.40 &   7.23   &   1.38 &   0.4248   &   0.0005 &   0.2837   &   0.0007   &   0.3822   &   0.0009   &   7.189   &   0.072   &   52.69   &   0.53   &   -16.87   &   -0.17   \\  2458195.77803   &   -12.79   &   0.54 &   6.33   &   7.12 &   7.06   &   1.23 &   0.4253   &   0.0005 &   0.2839   &   0.0007   &   0.3799   &   0.0009   &   7.192   &   0.072   &   52.69   &   0.53   &   -15.75   &   -0.16   \\  2458195.78406   &   -12.88   &   0.54 &   6.89   &   7.76 &   7.61   &   1.43 &   0.4269   &   0.0005 &   0.2843   &   0.0007   &   0.3825   &   0.0009   &   7.192   &   0.072   &   52.70   &   0.53   &   -18.31   &   -0.18   \\  2458197.76886   &   -11.48   &   0.58 &   7.76   &   6.50 &   7.05   &   1.29 &   0.4247   &   0.0006 &   0.2854   &   0.0007   &   0.3901   &   0.0009   &   7.193   &   0.072   &   52.69   &   0.53   &   -15.85   &   -0.16   \\  2458197.77518   &   -12.31   &   0.57 &   -2.83   &   6.89 &   6.40   &   1.13 &   0.4264   &   0.0006 &   0.2852   &   0.0007   &   0.3894   &   0.0009   &   7.192   &   0.072   &   52.69   &   0.53   &   -17.90   &   -0.18   \\  2458197.78127   &   -11.97   &   0.52 &   -1.04   &   7.24 &   7.44   &   1.24 &   0.4255   &   0.0005 &   0.2849   &   0.0006   &   0.3905   &   0.0008   &   7.194   &   0.072   &   52.69   &   0.53   &   -16.86   &   -0.17   \\  2458222.76220   &   18.17   &   0.70 &   1.52   &   7.62 &   5.31   &   1.58 &   0.4286   &   0.0007 &   0.2854   &   0.0009   &   0.3900   &   0.0011   &   7.191   &   0.072   &   52.71   &   0.53   &   -17.39   &   -0.17   \\  2458222.76882   &   15.31   &   0.77 &   0.24   &   9.41 &   3.66   &   1.86 &   0.4281   &   0.0008 &   0.2864   &   0.0010   &   0.3789   &   0.0013   &   7.193   &   0.072   &   52.70   &   0.53   &   -14.90   &   -0.15   \\  2458222.77465   &   14.30   &   0.66 &   -6.83   &   6.58 &   6.26   &   1.63 &   0.4278   &   0.0006 &   0.2863   &   0.0008   &   0.3803   &   0.0011   &   7.194   &   0.072   &   52.69   &   0.53   &   -13.67   &   -0.14   \\  2458223.77444   &   13.00   &   0.71 &   -1.28   &   6.88 &   5.74   &   1.58 &   0.4246   &   0.0007 &   0.2877   &   0.0009   &   0.3874   &   0.0012   &   7.192   &   0.072   &   52.70   &   0.53   &   -11.55   &   -0.12   \\  2458223.78055   &   13.22   &   0.72 &   -9.98   &   7.69 &   4.24   &   1.45 &   0.4243   &   0.0007 &   0.2858   &   0.0009   &   0.3870   &   0.0011   &   7.190   &   0.072   &   52.72   &   0.53   &   -17.95   &   -0.18   \\  2458223.78682   &   12.74   &   0.87 &   4.94   &   9.28 &   6.03   &   1.90 &   0.4237   &   0.0008 &   0.2857   &   0.0011   &   0.3879   &   0.0014   &   7.196   &   0.072   &   52.69   &   0.53   &   -14.50   &   -0.15   \\  2458224.70939   &   10.84   &   0.78 &   4.17   &   8.26 &   4.19   &   1.76 &   0.4296   &   0.0008 &   0.2842   &   0.0011   &   0.3811   &   0.0014   &   7.196   &   0.072   &   52.70   &   0.53   &   -14.53   &   -0.15   \\  2458224.71561   &   12.62   &   0.59 &   -4.39   &   7.02 &   7.39   &   1.41 &   0.4285   &   0.0006 &   0.2869   &   0.0007   &   0.3794   &   0.0010   &   7.192   &   0.072   &   52.68   &   0.53   &   -14.51   &   -0.15   \\  2458224.72177   &   12.92   &   0.55 &   -5.42   &   7.61 &   8.58   &   1.11 &   0.4282   &   0.0005 &   0.2839   &   0.0007   &   0.3764   &   0.0009   &   7.194   &   0.072   &   52.68   &   0.53   &   -15.94   &   -0.16   \\  2458225.69613   &   12.07   &   0.98 &   -11.31   &   8.81 &   2.48   &   2.03 &   0.4232   &   0.0011 &   0.2970   &   0.0014   &   0.3803   &   0.0017   &   7.191   &   0.072   &   52.71   &   0.53   &   -13.67   &   -0.14   \\  2458225.70234   &   9.99   &   1.06 &   0.22   &   10.09 &   5.05   &   2.30 &   0.4200   &   0.0012 &   0.2955   &   0.0015   &   0.3778   &   0.0019   &   7.192   &   0.072   &   52.71   &   0.53   &   -11.61   &   -0.12   \\  2458225.70850   &   12.25   &   1.02 &   0.14   &   9.39 &   3.27   &   2.35 &   0.4209   &   0.0011 &   0.2864   &   0.0014   &   0.3786   &   0.0018   &   7.188   &   0.072   &   52.72   &   0.53   &   -12.72   &   -0.13   \\  2458249.46316   &   -3.42   &   0.74 &   3.13   &   7.03 &   6.24   &   1.30 &   0.4246   &   0.0008 &   0.2941   &   0.0010   &   0.3796   &   0.0013   &   7.193   &   0.072   &   52.70   &   0.53   &   -16.45   &   -0.16   \\  2458249.46927   &   -2.71   &   0.75 &   -2.77   &   7.58 &   2.76   &   1.57 &   0.4234   &   0.0008 &   0.2939   &   0.0010   &   0.3781   &   0.0013   &   7.193   &   0.072   &   52.73   &   0.53   &   -15.06   &   -0.15   \\  2458249.47548   &   -2.56   &   0.60 &   -0.06   &   6.03 &   3.47   &   1.29 &   0.4241   &   0.0007 &   0.2847   &   0.0008   &   0.3807   &   0.0011   &   7.191   &   0.072   &   52.71   &   0.53   &   -12.57   &   -0.13   \\  2458516.82380   &   6.44   &   0.67 &   0.23   &   7.83 &   3.90   &   1.57 &   0.4275   &   0.0008 &   0.2841   &   0.0009   &   0.4024   &   0.0012   &   7.187   &   0.072   &   52.73   &   0.53   &   -18.82   &   -0.19   \\
  
\hline         

\end{tabular}}

\end{table*}
\begin{table*}
\caption{HARPS Doppler measurements and activity index measurements of HD\,107148, derived with SERVAL (Cont.)} 
\label{tab:HARPS_2}

\centering  
\resizebox{\textwidth}{!}{
\begin{tabular}{c c c c c c c c c c c c c c c c c c c} 

\hline\hline    
\noalign{\vskip 0.5mm}

Epoch & RV & $\sigma_{RV}$  &  CRX & $\sigma_{\rm CRX}$ &  dLW & $\sigma_{\rm dLW} $  &  H$\alpha$ & $\sigma_{H\alpha}$  &  NaD$_1$ & $\sigma_{\rm NaD_1}$  &  NaD$_2$ & $\sigma_{\rm NaD_2}$  &  FWHM & $\sigma_{\rm FWHM} $  &  Cont & $\sigma_{\rm Cont}$  &  BIS & $\sigma_{\rm BIS} $  \\ 

[JD] & [m\,s$^{-1}$] &  [m\,s$^{-1}$]  &  [m\,s$^{-1}$] &  [m\,s$^{-1}$]  &  [m\,s$^{-1}$ * km\,s$^{-1}$]& [m\,s$^{-1}$ * km\,s$^{-1}$]  &   &   &  &   &   &   &  [km\,s$^{-1}$]& [km\,s$^{-1}$]  &   &   &   [km\,s$^{-1}$]&  k[m\,s$^{-1}$]  \\  

\hline     
\noalign{\vskip 0.5mm}   

2458516.83563   &   7.31   &   0.66 &   -5.71   &   8.04 &   3.34   &   1.15 &   0.4277   &   0.0008 &   0.2807   &   0.0009   &   0.4010   &   0.0012   &   7.184   &   0.072   &   52.74   &   0.53   &   -19.10   &   -0.19   \\  2458517.85673   &   6.23   &   0.62 &   14.21   &   6.42 &   6.51   &   1.35 &   0.4282   &   0.0007 &   0.2787   &   0.0009   &   0.3978   &   0.0012   &   7.189   &   0.072   &   52.74   &   0.53   &   -17.09   &   -0.17   \\  2458517.86218   &   5.00   &   0.70 &   11.41   &   6.65 &   4.17   &   1.58 &   0.4290   &   0.0008 &   0.2782   &   0.0009   &   0.3977   &   0.0012   &   7.189   &   0.072   &   52.74   &   0.53   &   -15.76   &   -0.16   \\  2458528.86119   &   5.90   &   0.61 &   1.48   &   7.94 &   2.10   &   1.10 &   0.4286   &   0.0007 &   0.2855   &   0.0008   &   0.3927   &   0.0011   &   7.186   &   0.072   &   52.75   &   0.53   &   -15.43   &   -0.15   \\  2458528.86740   &   5.70   &   0.65 &   5.32   &   9.44 &   3.80   &   1.56 &   0.4289   &   0.0007 &   0.2854   &   0.0009   &   0.3928   &   0.0011   &   7.186   &   0.072   &   52.75   &   0.53   &   -15.47   &   -0.15   \\  2458528.87373   &   5.65   &   0.64 &   11.84   &   8.32 &   2.27   &   1.28 &   0.4283   &   0.0007 &   0.2861   &   0.0009   &   0.3932   &   0.0011   &   7.189   &   0.072   &   52.73   &   0.53   &   -16.75   &   -0.17   \\  2458530.89322   &   6.91   &   0.52 &   -9.43   &   5.50 &   2.14   &   1.10 &   0.4281   &   0.0005 &   0.2849   &   0.0007   &   0.3953   &   0.0009   &   7.183   &   0.072   &   52.76   &   0.53   &   -18.91   &   -0.19   \\  2458530.89942   &   6.64   &   0.53 &   1.52   &   6.29 &   2.97   &   1.26 &   0.4283   &   0.0005 &   0.2842   &   0.0007   &   0.3962   &   0.0009   &   7.189   &   0.072   &   52.74   &   0.53   &   -17.48   &   -0.17   \\  2458530.90556   &   5.29   &   0.55 &   0.66   &   7.16 &   5.03   &   1.18 &   0.4285   &   0.0006 &   0.2843   &   0.0007   &   0.3962   &   0.0009   &   7.187   &   0.072   &   52.72   &   0.53   &   -16.99   &   -0.17   \\  2458538.88884   &   9.17   &   0.55 &   0.88   &   6.25 &   3.92   &   1.35 &   0.4274   &   0.0006 &   0.2881   &   0.0007   &   0.3940   &   0.0009   &   7.187   &   0.072   &   52.73   &   0.53   &   -13.79   &   -0.14   \\  2458538.89512   &   9.60   &   0.58 &   3.08   &   7.81 &   2.33   &   1.11 &   0.4270   &   0.0006 &   0.2894   &   0.0007   &   0.3928   &   0.0009   &   7.183   &   0.072   &   52.74   &   0.53   &   -16.25   &   -0.16   \\  2458538.90116   &   9.17   &   0.60 &   13.62   &   7.83 &   1.35   &   1.13 &   0.4272   &   0.0006 &   0.2884   &   0.0008   &   0.3924   &   0.0010   &   7.183   &   0.072   &   52.76   &   0.53   &   -17.07   &   -0.17   \\  2458542.85771   &   4.86   &   0.57 &   -12.12   &   7.94 &   2.09   &   1.28 &   0.4293   &   0.0006 &   0.2825   &   0.0007   &   0.3909   &   0.0009   &   7.187   &   0.072   &   52.75   &   0.53   &   -17.44   &   -0.17   \\  2458542.86380   &   4.89   &   0.55 &   -1.44   &   5.98 &   2.15   &   1.23 &   0.4290   &   0.0006 &   0.2825   &   0.0007   &   0.3908   &   0.0009   &   7.193   &   0.072   &   52.74   &   0.53   &   -16.93   &   -0.17   \\  2458542.86967   &   5.32   &   0.61 &   -4.68   &   7.43 &   0.20   &   1.31 &   0.4291   &   0.0006 &   0.2821   &   0.0008   &   0.3918   &   0.0010   &   7.184   &   0.072   &   52.75   &   0.53   &   -19.44   &   -0.19   \\  2458546.84827   &   4.11   &   0.65 &   10.39   &   8.77 &   -0.47   &   1.46 &   0.4270   &   0.0007 &   0.2815   &   0.0008   &   0.3780   &   0.0011   &   7.188   &   0.072   &   52.76   &   0.53   &   -15.24   &   -0.15   \\  2458546.85454   &   3.19   &   0.64 &   9.17   &   9.17 &   1.06   &   1.38 &   0.4304   &   0.0007 &   0.2820   &   0.0008   &   0.3805   &   0.0011   &   7.188   &   0.072   &   52.75   &   0.53   &   -22.55   &   -0.23   \\  2458546.85995   &   2.53   &   0.94 &   6.72   &   10.57 &   2.74   &   1.81 &   0.4268   &   0.0012 &   0.2815   &   0.0013   &   0.3828   &   0.0017   &   7.190   &   0.072   &   52.71   &   0.53   &   -14.24   &   -0.14   \\  2458549.83603   &   7.59   &   0.74 &   16.37   &   7.77 &   2.73   &   1.62 &   0.4268   &   0.0008 &   0.2822   &   0.0010   &   0.3896   &   0.0013   &   7.189   &   0.072   &   52.74   &   0.53   &   -16.50   &   -0.16   \\  2458549.84225   &   7.30   &   0.74 &   -1.73   &   9.74 &   1.57   &   1.59 &   0.4297   &   0.0008 &   0.2819   &   0.0010   &   0.3787   &   0.0012   &   7.187   &   0.072   &   52.77   &   0.53   &   -14.80   &   -0.15   \\  2458549.84834   &   6.93   &   0.70 &   -6.41   &   10.41 &   0.43   &   1.22 &   0.4272   &   0.0008 &   0.2836   &   0.0009   &   0.3796   &   0.0012   &   7.190   &   0.072   &   52.76   &   0.53   &   -19.08   &   -0.19   \\  2458565.78695   &   -9.61   &   0.54 &   -4.12   &   7.80 &   0.80   &   1.04 &   0.4249   &   0.0005 &   0.2841   &   0.0006   &   0.3897   &   0.0008   &   7.186   &   0.072   &   52.74   &   0.53   &   -18.58   &   -0.19   \\  2458565.79317   &   -10.17   &   0.52 &   6.53   &   6.98 &   0.33   &   1.31 &   0.4254   &   0.0005 &   0.2834   &   0.0006   &   0.3889   &   0.0008   &   7.187   &   0.072   &   52.74   &   0.53   &   -19.36   &   -0.19   \\  2458565.79932   &   -9.07   &   0.54 &   -5.42   &   7.69 &   2.69   &   1.32 &   0.4242   &   0.0005 &   0.2842   &   0.0006   &   0.3904   &   0.0008   &   7.188   &   0.072   &   52.73   &   0.53   &   -18.39   &   -0.18   \\  2458566.74595   &   -6.86   &   0.52 &   -0.31   &   10.74 &   4.42   &   1.19 &   0.4272   &   0.0005 &   0.2911   &   0.0006   &   0.3906   &   0.0008   &   7.190   &   0.072   &   52.73   &   0.53   &   -18.76   &   -0.19   \\  2458566.75240   &   -7.37   &   0.58 &   -1.78   &   9.82 &   2.28   &   1.34 &   0.4255   &   0.0006 &   0.2918   &   0.0007   &   0.3904   &   0.0009   &   7.194   &   0.072   &   52.73   &   0.53   &   -18.55   &   -0.19   \\  2458566.75832   &   -8.57   &   0.55 &   5.99   &   10.45 &   1.07   &   1.19 &   0.4270   &   0.0006 &   0.2904   &   0.0007   &   0.3903   &   0.0009   &   7.188   &   0.072   &   52.74   &   0.53   &   -16.24   &   -0.16   \\  2458567.82953   &   -6.61   &   0.64 &   6.09   &   8.03 &   1.89   &   1.15 &   0.4244   &   0.0006 &   0.2911   &   0.0008   &   0.3898   &   0.0010   &   7.191   &   0.072   &   52.72   &   0.53   &   -19.40   &   -0.19   \\  2458567.83575   &   -5.46   &   0.68 &   -3.92   &   8.30 &   1.88   &   1.41 &   0.4227   &   0.0007 &   0.2916   &   0.0008   &   0.3887   &   0.0010   &   7.189   &   0.072   &   52.72   &   0.53   &   -17.08   &   -0.17   \\  2458567.84185   &   -7.09   &   0.69 &   10.03   &   8.36 &   3.84   &   1.37 &   0.4235   &   0.0007 &   0.2907   &   0.0008   &   0.3920   &   0.0011   &   7.191   &   0.072   &   52.73   &   0.53   &   -17.86   &   -0.18   \\  2458586.72225   &   -7.99   &   0.54 &   10.23   &   6.23 &   0.40   &   1.18 &   0.4240   &   0.0006 &   0.2854   &   0.0007   &   0.3797   &   0.0008   &   7.187   &   0.072   &   52.74   &   0.53   &   -16.84   &   -0.17   \\  2458600.73936   &   1.78   &   0.64 &   -10.72   &   7.82 &   -0.55   &   1.37 &   0.4248   &   0.0006 &   0.2957   &   0.0008   &   0.3827   &   0.0010   &   7.190   &   0.072   &   52.74   &   0.53   &   -19.00   &   -0.19   \\  2458600.74609   &   0.62   &   0.83 &   -7.70   &   8.62 &   -2.21   &   1.55 &   0.4236   &   0.0008 &   0.2947   &   0.0010   &   0.3827   &   0.0013   &   7.188   &   0.072   &   52.75   &   0.53   &   -17.95   &   -0.18   \\  2458600.75161   &   2.56   &   0.72 &   11.07   &   7.30 &   -3.27   &   1.41 &   0.4246   &   0.0007 &   0.2938   &   0.0009   &   0.3827   &   0.0011   &   7.193   &   0.072   &   52.75   &   0.53   &   -19.19   &   -0.19   \\  2458608.61031   &   11.69   &   0.70 &   -4.08   &   6.69 &   -1.13   &   1.47 &   0.4291   &   0.0008 &   0.2936   &   0.0009   &   0.3867   &   0.0012   &   7.192   &   0.072   &   52.74   &   0.53   &   -17.45   &   -0.17   \\  2458608.61647   &   11.53   &   0.68 &   3.95   &   7.00 &   0.82   &   1.20 &   0.4259   &   0.0007 &   0.2946   &   0.0009   &   0.3891   &   0.0011   &   7.190   &   0.072   &   52.74   &   0.53   &   -19.55   &   -0.20   \\  2458608.62251   &   10.39   &   0.68 &   11.81   &   7.85 &   -2.87   &   1.37 &   0.4265   &   0.0008 &   0.2937   &   0.0009   &   0.3875   &   0.0011   &   7.189   &   0.072   &   52.75   &   0.53   &   -14.89   &   -0.15   \\  2458609.58060   &   11.05   &   0.73 &   6.19   &   7.82 &   -0.65   &   1.33 &   0.4251   &   0.0008 &   0.2928   &   0.0009   &   0.3856   &   0.0012   &   7.194   &   0.072   &   52.74   &   0.53   &   -20.46   &   -0.20   \\  2458609.58704   &   10.61   &   0.68 &   11.78   &   7.85 &   0.39   &   1.31 &   0.4251   &   0.0007 &   0.2928   &   0.0009   &   0.3883   &   0.0011   &   7.191   &   0.072   &   52.74   &   0.53   &   -17.48   &   -0.17   \\  2458609.59291   &   11.63   &   0.70 &   -3.27   &   7.34 &   -1.78   &   1.61 &   0.4241   &   0.0008 &   0.2946   &   0.0009   &   0.3883   &   0.0011   &   7.190   &   0.072   &   52.75   &   0.53   &   -14.06   &   -0.14   \\  2458627.66143   &   10.13   &   0.97 &   -10.35   &   11.99 &   -3.93   &   2.08 &   0.4252   &   0.0011 &   0.2912   &   0.0013   &   0.3824   &   0.0016   &   7.184   &   0.072   &   52.76   &   0.53   &   -20.12   &   -0.20   \\  2458627.66759   &   9.64   &   0.96 &   -0.22   &   9.40 &   -0.85   &   1.88 &   0.4270   &   0.0011 &   0.2899   &   0.0012   &   0.3805   &   0.0016   &   7.183   &   0.072   &   52.75   &   0.53   &   -16.45   &   -0.16   \\  2458627.67363   &   11.91   &   1.01 &   4.34   &   11.17 &   2.58   &   1.99 &   0.4262   &   0.0011 &   0.2935   &   0.0013   &   0.3792   &   0.0017   &   7.188   &   0.072   &   52.72   &   0.53   &   -20.22   &   -0.20   \\  2458667.50998   &   -2.86   &   0.64 &   17.00   &   8.56 &   -1.48   &   1.42 &   0.4311   &   0.0006 &   0.2916   &   0.0008   &   0.3868   &   0.0010   &   7.188   &   0.072   &   52.74   &   0.53   &   -19.73   &   -0.20   \\  2458667.51614   &   -1.58   &   0.67 &   6.64   &   7.82 &   -0.63   &   1.49 &   0.4291   &   0.0007 &   0.2916   &   0.0008   &   0.3872   &   0.0010   &   7.188   &   0.072   &   52.75   &   0.53   &   -18.88   &   -0.19   \\  2458668.48940   &   -2.84   &   0.96 &   -5.46   &   9.95 &   -4.90   &   1.96 &   0.4272   &   0.0010 &   0.2929   &   0.0012   &   0.3793   &   0.0016   &   7.187   &   0.072   &   52.76   &   0.53   &   -20.89   &   -0.21   \\  2458668.49525   &   -3.53   &   0.91 &   -12.44   &   9.14 &   -2.30   &   1.77 &   0.4246   &   0.0010 &   0.2919   &   0.0012   &   0.3803   &   0.0015   &   7.190   &   0.072   &   52.75   &   0.53   &   -20.41   &   -0.20   \\  2458668.50151   &   -1.27   &   0.91 &   3.15   &   9.74 &   -3.88   &   1.98 &   0.4302   &   0.0010 &   0.2920   &   0.0012   &   0.3795   &   0.0015   &   7.187   &   0.072   &   52.75   &   0.53   &   -17.14   &   -0.17   \\
  
\hline       
\end{tabular}}

\end{table*}
\begin{table*}
\caption{CARMENES Doppler measurements and activity index measurements of HD\,107148 in visible light.} 
\label{tab:CARM_VIS} 

\centering  
\resizebox{\textwidth}{!}{
\begin{tabular}{c c c c c c c c c c c c c} 

\hline\hline    
\noalign{\vskip 0.5mm}

Epoch & RV & $\sigma_{RV}$  &  CRX & $\sigma_{\rm CRX}$ &  dLW & $\sigma_{\rm dLW} $  &  H$\alpha$ & $\sigma_{H\alpha}$  &  NaD$_1$ & $\sigma_{\rm NaD_1}$  &  NaD$_2$ & $\sigma_{\rm NaD_2}$ \\  

[JD] & [m\,s$^{-1}$] &  [m\,s$^{-1}$]  &  [m\,s$^{-1}$] &  [m\,s$^{-1}$]  &  [m\,s$^{-1}$ * km\,s$^{-1}$]& [m\,s$^{-1}$ * km\,s$^{-1}$]  &   &   &  &   &   &  \\

\hline     
\noalign{\vskip 0.5mm}

2457768.71507   &   4.81   &   2.14 &   -7.07   &   16.03 &   38.79   &   3.51 &   0.4157   &   0.0009 &   0.2778   &   0.0014   &   0.3357   &   0.0016   \\  2457768.72281   &   0.09   &   2.26 &   -17.39   &   17.49 &   48.24   &   3.72 &   0.4148   &   0.0009 &   0.2899   &   0.0013   &   0.3470   &   0.0015   \\  2457813.68579   &   -17.68   &   1.29 &   -16.40   &   9.73 &   -15.96   &   1.95 &   0.4147   &   0.0006 &   0.2825   &   0.0009   &   0.3568   &   0.0010   \\  2457813.69267   &   -19.87   &   1.31 &   -0.74   &   10.60 &   -10.83   &   1.94 &   0.4158   &   0.0005 &   0.2803   &   0.0008   &   0.3574   &   0.0009   \\  2457813.69604   &   -22.99   &   1.16 &   8.53   &   9.16 &   -14.68   &   2.22 &   0.4164   &   0.0005 &   0.2804   &   0.0008   &   0.3563   &   0.0009   \\  2457813.69944   &   -17.05   &   1.31 &   2.09   &   10.39 &   -13.12   &   2.22 &   0.4162   &   0.0006 &   0.2805   &   0.0008   &   0.3539   &   0.0010   \\  2457813.70282   &   -19.25   &   1.26 &   -9.63   &   10.07 &   -12.95   &   2.32 &   0.4167   &   0.0005 &   0.2800   &   0.0008   &   0.3566   &   0.0009   \\  2457854.54602   &   2.99   &   2.55 &   -1.49   &   20.39 &   5.50   &   4.57 &   0.4094   &   0.0010 &   0.2658   &   0.0014   &   0.3241   &   0.0016   \\  2457854.55834   &   -2.69   &   1.87 &   -11.98   &   15.08 &   16.79   &   2.85 &   0.4118   &   0.0007 &   0.2834   &   0.0010   &   0.3420   &   0.0011   \\  2457854.56173   &   4.45   &   1.80 &   16.48   &   14.38 &   15.01   &   2.13 &   0.4122   &   0.0007 &   0.2835   &   0.0010   &   0.3459   &   0.0012   \\  2457854.56512   &   2.05   &   1.78 &   11.42   &   14.41 &   18.96   &   2.42 &   0.4122   &   0.0007 &   0.2855   &   0.0010   &   0.3459   &   0.0011   \\  2457854.56849   &   1.20   &   2.06 &   3.69   &   16.60 &   13.49   &   3.13 &   0.4130   &   0.0009 &   0.2842   &   0.0013   &   0.3464   &   0.0014   \\

\hline           
\end{tabular}}

\end{table*}
\begin{table*}
\caption{CARMENES Doppler measurements and activity index measurements of HD\,107148 in near infrared.} 
\label{tab:CARM_NIR} 

\centering  
\resizebox{\textwidth}{!}{
\begin{tabular}{c c c c c c c} 

\hline\hline    
\noalign{\vskip 0.5mm}

Epoch & RV & $\sigma_{RV}$  &  CRX & $\sigma_{\rm CRX}$ &  dLW & $\sigma_{\rm dLW} $ \\  

[JD] & [m\,s$^{-1}$] &  [m\,s$^{-1}$]  &  [m\,s$^{-1}$] &  [m\,s$^{-1}$]  &  [m\,s$^{-1}$ * km\,s$^{-1}$]& [m\,s$^{-1}$ * km\,s$^{-1}$]  \\ 

\hline     
\noalign{\vskip 0.5mm}

2457768.71526   &   59.26   &   6.26 &   -5.85   &   33.49 &   41.62   &   5.58   \\  2457768.72282   &   52.97   &   7.23 &   -4.46   &   39.66 &   32.96   &   4.06   \\  2457813.68586   &   87.84   &   6.15 &   12.92   &   32.77 &   -4.81   &   5.33   \\  2457813.69181   &   69.60   &   8.26 &   93.76   &   52.85 &   -27.71   &   7.24   \\  2457813.69694   &   57.79   &   7.65 &   35.20   &   47.86 &   -22.46   &   4.91   \\  2457813.70033   &   55.64   &   7.73 &   65.39   &   48.79 &   -18.56   &   6.25   \\  2457813.70334   &   66.94   &   7.21 &   42.93   &   45.81 &   -30.15   &   6.36   \\  2457854.54653   &   47.69   &   4.50 &   -18.00   &   24.45 &   22.68   &   3.99   \\  2457854.55834   &   41.86   &   4.99 &   -1.89   &   28.18 &   6.96   &   3.71   \\  2457854.56224   &   167.85   &   10.75 &   146.02   &   56.56 &   54.93   &   3.90   \\  2457854.56613   &   38.49   &   5.07 &   -18.72   &   28.01 &   12.72   &   3.96   \\  2457854.57002   &   44.64   &   4.83 &   -27.94   &   26.29 &   11.97   &   3.89   \\

\hline           
\end{tabular}}

\end{table*}

\end{appendix}

\end{document}